# FROM THE BERLIN "ENTWURF" FIELD EQUATIONS TO THE EINSTEIN TENSOR II: November 1915 until March 1916

Galina Weinstein

*Visitor Scholar, Center for Einstein Studies, Philosophy Department, Boston University*

*January 25, 2012*

I discuss Einstein's path-breaking November 1915 General Relativity papers. I show that Einstein's field equations of November 25, 1915 with an additional term on the right hand side involving the trace of the energy-momentum tensor appear to have sprung from his first November 1915 paper: the November 4, 1915 equations. Second paper among three papers.

## 1 Introduction: David Hilbert Enters the Game, the priority dispute in a nutshell

According to Stachel Einstein's Odyssey to general covariance during November 1915 went through two stages:[1]

1) Sometime *in October 1915* Einstein dropped the Einstein-Grossman "Entwurf" theory. *During October 1915* Einstein realized that the key to the solution lays in equations (14) and (17) of page 1041 of his 1914 review article "The Formal Foundation of the General Theory of Relativity", and adopted the determinant in equation (14), $\sqrt{-g} = 1$ as a *postulate*.[2] This led him to general covariance. *Starting on November 4* Einstein gradually expanded the range of the covariance of his field equations.

2) *Between November 4 and November 11* Einstein realized that he did not need this postulate and he adopted it as a coordinate condition to simplify the field equations. Einstein was able to write the field equations of gravitation in a general covariant form. In the November 11 field equations the trace of the energy-momentum tensor vanishes. *Between November 18 and November 25* Einstein found that he could write the field equations with an additional term on the right hand side of the field equations involving the trace of the energy-momentum tensor, which now need not vanish. These were the November 25 field equations.

Let us start with the first stage. During October 1915 Einstein presumably started with page 1041 of the 1914 paper, and then moved on to the next page 1042, and found a problem with the "*fundamental tensor of Ricci and Levi-Civita*", equation (19), which he *did not use* in his field equations of 1914.[3] In his 1914 paper Einstein had to manipulate a derivation in order to demonstrate the tensorial character of $G_{ik}^{lm}$; the problem was the $\frac{1}{\sqrt{g}}$ and the $\sqrt{g}$ factors that entered into equations (21) and (19), and thus led to $G_{iklm} \neq G^{iklm}$ .[4] His field equations and generally the equations of his 1914 paper made a distinction between tensors and what he called "V-tensors" (tensor densities). The V-tensors were required because of the factors $\sqrt{-g}$ and $1/\sqrt{-g}$ which were not equal to 1. This complicated the equations of the Einstein-Grossmann theory. By November 4, 1915 Einstein realized that when only substitutions of determinant 1 are permitted this simplified the equations as the above factor was omitted, and $G_{iklm} = G^{iklm}$.

Einstein flipped a few extra pages of his 1914 paper, and arrived at page 1053. On page 1053 Einstein wrote the Riemann-Christoffel Tensor.[5] In the first paper of November 4, 1915 Einstein wrote about the Ricci tensor $G_{im}$ that the derivation of this tensor was better obtained from a different form of the Riemann-Christoffel tensor (ik, lm). In a footnote Einstein explained that he had given this proof already in his 1914 paper on page 1053.[6] Einstein defined the second term in equation (13), $R_{im}$, like he had done three years earlier on page 22R of the Zurich Notebook.[7] According to the footnote, page 1053 of the 1914 paper was very likely a trigger for Einstein to come back to his calculations from three years earlier.

Einstein started to write the November 4 paper and adopted the postulate that his field equations were covariant with respect to arbitrary transformations of a determinant equal to 1.[8]

While he was writing the November 4 paper, Einstein realized that by contracting the Ricci tensor one would arrive at a scalar equation [(21a)] that implied that it was impossible to choose a coordinate system in which $\sqrt{-g} = 1$. If we choose such a coordinate system, then log 1 = 0 and the scalar of the energy tensor $\sum_\sigma T_\sigma^\sigma$ is set to zero.[9] Nevertheless, Einstein published this result in the November 4 paper and mentioned the problem with this result.

Renn and Stachel explain that, "the physical meaning of condition" (21a) "was entirely obscure. It was therefore incumbent upon Einstein to find a physical interpretation of it or to modify his theory once more in order to get rid of it. He soon succeeded in doing both, and formulated his new view in an addendum to the first note, published on November 11".[10]

On November 7, 1915, Einstein sent David Hilbert the proofs to his first paper of November 4, and he wanted Hilbert to look at this work. Hilbert *also read Einstein's 1914 paper*, and found some mistake in this paper; Einstein wrote that his colleague Sommerfeld wrote him that Hilbert had objected to the 1914 "Entwurf" foundations paper. [11] The mistake is discussed later.

By November 10, 1915 Hilbert probably answered Einstein's letter, telling him about his system of electromagnetic theory of matter, the unified theory of gravitation and electromagnetism, in which the source of the gravitational field is the electromagnetic field. Hilbert goal was to develop an electromagnetic theory of matter, which would explain the stability of the electron.[12]

Between November 4 and November 11 it seems that Einstein was influence by Hilbert's physical attitude towards a field theory of matter. In his addendum to the first note, published on November 11 Einstein directly referred to the supporters of the electrodynamic worldview, "One now has to remember that, in accord with our knowledge, 'matter' is not to be conceived as something primitively given, or physically simple. There even are those, and not just a few, who hope to be able to reduce matter to purely electrodynamic processes, which of course would have to be done in a theory more complete than Maxwell's electrodynamics". Einstein probably discussed the electrodynamic worldview with Hilbert and felt that he was now in competition with the latter.[13]

In the addendum to the November 4 paper, the November 11 paper, Einstein added the following coordinate condition: "*we assume in the following that the condition* $\sum T_\mu^\mu = 0$ *is in general actually fulfilled* ".[14] This allowed Einstein to take the last step and to write the field equations of gravitation in a general covariant form.[15] He then dropped his November 4 postulate and adopted it as a coordinate condition, $\sqrt{-g} = 1$.

The day afterwards Einstein wrote Hilbert again. [16] He told him about the progress in his work. Hilbert replied and invited Einstein to come to Göttingen. [17] Hilbert explained to Einstein the main points of his new unified theory of gravitation and electromagnetism, and told Einstein that he had already discussed his discovery with Sommerfeld. He wanted next to explain it to Einstein. He thus invited him to come to hear his talk on November 16. Hilbert told Einstein that the latter's November 4 paper was entirely different from his own work.

With hindsight Hilbert's work was different from Einstein's November 4 paper in that, Hilbert eventually endeavored to derive generally covariant field equations for the combined gravitational and electromagnetic fields without explicitly writing down these equations. Hilbert accepted Einstein's 1914 Hole Argument against general covariance (after Einstein had silently dropped it). Hilbert was thus finally obliged to supplement his generally covariant field equations by four non-generally covariant field equations based on rather dubious energy considerations, which Hilbert would eventually drop later when he would publish his paper (after Einstein presented his final form of field equations to the Prussian Academy on November 25).[18]

Einstein replied and told Hilbert he could not come, but requested a copy of his work.[19] In response, Hilbert perhaps sent a copy of the lecture he had given on the subject on November 16, or else a copy of a manuscript of the paper he would present five days later on November 20 to the Royal Society in Göttingen. [20]

Einstein was already less patient after he had received Hilbert's work. He replied to Hilbert on November 18 telling him that his work agrees – as far as he could see – exactly with what he had found in the last few weeks and have already presented to the Prussian Academy.[21] Einstein was in competition with Hilbert and appeared to have been still influenced by his unified theory of matter, gravitation and electromagnetism until November 18.[22] Indeed on Thursday, November 18, Einstein presented to the Prussian Academy his solution to the longstanding problem of the precession of the perihelion of Mercury, on the basis of his *November 11 General theory of relativity*.

The day afterwards Hilbert sent a polite letter in which he congratulated Einstein on overcoming the perihelion motion. He was quite astonished that Einstein calculated so rapidly the precession of Mercury's perihelion.[23] In fact the basic calculation has already been done two years earlier with Besso in the Einstein-Besso manuscript. Einstein transferred the basic framework of the calculation from the Einstein-Besso manuscript, and corrected it according to his November field equations. [24]

In the 1913 calculation of the Einstein-Besso manuscript, the gravitational field (of the "Entwurf" theory) was represented by $g_{\mu\nu}$, and the perihelion advance of Mercury was calculated by first and second approximations of $g_{\mu\nu}$. In 1915 the gravitational field was represented by $\Gamma^{\tau}_{\rho\sigma}$ , and Einstein used the components $\Gamma^{\tau}_{\rho\sigma}$ of the

gravitational field of the sun in order to find the perihelion advance of Mercury.[25]

On page 1 of the Einstein-Besso manuscript Einstein first took the 44 component of the "Entwurf" field equation ["Entwurf" contravarient and covariant field equations respectively, $\Delta(\gamma) = \chi(\Theta_{\mu\nu} + \vartheta_{\mu\nu})$, $-D_{\mu\nu}(g) = \chi(t_{\mu\nu} + T_{\mu\nu})$]: [26]

$$-\Delta\gamma_{44} = \frac{\kappa\rho_0}{c^2{}_0},$$

where, $\Delta$ is the Laplacian, $\kappa$ is the gravitational constant: $\kappa = K(8\pi/c^2{}_0)$, $\rho_0$ is the mass density of the sun, and $c_0$ is the speed of light in vacuum.

The solution to this equation gives the contravariant metric field $\gamma_{44}$ of the sun to first order:

$$\frac{1}{c^2{}_0}\left(1+\frac{A}{r}\right).$$

And the covariant metric field $g_{44}$ of the sun to first order:

$$c^2{}_0\left(1-\frac{A}{r}\right).$$[27]

Next, the first order solutions are substituted in the "Entwurf" equations and second order contributions are evaluated when one assumes that the field of the sun is spherically symmetric:

$$\gamma_{44}=-\frac{1}{c^2{}_0}\left(1+\frac{A}{r}+\frac{5}{8}\frac{A^2}{r^2}\right),\; g_{44}=c^2{}_0\left(1-\frac{A}{r}+\frac{3}{8}\frac{A^2}{r^2}\right).$$[28]

The next step is equations of motion (not yet geodesic equation) for a point mass moving in the (weak-field) metric field of a static sun to second order (Euler-Lagrange equations for the action:

$S = \int H\, dt$, with the Lagrangian $H = - m\, ds/dt$).[29]

The Hamiltonian for a unit mass-point in the field of a static sun is the following:

$$E=-g_{44}\frac{dt}{ds}.$$

Besso derived the angular momentum conservation, which he called "the area law" (Flächensatz):

$$\frac{d}{dt}=\frac{y\dot{x}-x\dot{y}}{\frac{ds}{dt}}=o.$$

From this follows Kepler's second law. According to the "Entwurf" theory the relation between angular momentum conservation and Kepler's second law is more complicated. However, Besso still referred to the above equation as Kepler's law and called it "the area law".[30] Besso then defined the "Flachensatzkonstante" $Bc_0$, and "Flachengeschwindigkeit" $\dot{f}$ and wrote:

$$2\dot{f} = y\dot{x} - x\dot{y} = B\frac{ds}{dt}.$$ [31]

On page 9 Besso wrote an equation expressing conservation of angular momentum and energy:

$$2\dot{f} = \varphi\dot{r}^2 = BW, \;\; E = \frac{g_{44}}{W}. \quad W = \frac{ds}{dt}.$$

Inserting the metric field of a static sun to second order into the equation of $2\dot{f}$ and E above, and after rearrangements and additional manipulations Besso finally arrived at an equation on the bottom of page 9:

$$d\varphi = \frac{\frac{F}{c_0}\left(1-\frac{A}{r}\right)dr}{\sqrt{-\varepsilon r^4 + (1+2\varepsilon)\cdot A\cdot r^3 - \left(\frac{11}{8}A^2 + \frac{F^2}{c^2{}_0}\right)r^2 + 2\frac{F^2}{c^2{}_0}A^2\cdot r}} dr.$$ [32]

To find the advance of the perihelion, dφ has to be integrated between the values of r at perihelion and aphelion.[33] After some calculations and rearrangements Besso arrived at this result on page 14:

$$\int d\varphi = \pi\left(1 + \frac{5}{8}\frac{A}{a(1-e^2)}\right),$$

where, a is the semi-major axis and e the eccentricity of the elliptical orbit.[34]

From this equation it follows that according to the "Entwurf" theory the field of a static sun produces an advance of the perihelion of:

$$\pi\frac{5}{4}\frac{A}{a(1-e^2)}$$

per revolution.[35]

In November 1915 Einstein could calculate so rapidly the precession of Mercury's perihelion for another reason. Einstein's November 11 field equations for the metric tensor are the field equations for the gravitational field in the November 18 paper. The condition $\sqrt{-g} = 1$, implied by the assumption of an electromagnetic origin of

matter, was essential for Einstein's calculation of the precession of Mercury's perihelion.[36]

The November 11 field equations are non-linear partial differential equations of the second rank, and there is no general solution to these equations. Solving the field equations give the components of the metric tensor $g_{\mu\nu}$. In his November 18 paper Einstein tried to find approximate solutions.

Einstein was looking for the equation of a point moving along the geodesic line in the gravitational field of the sun. The solar system may be looked upon as an isolated mass, which is far away from other masses in the universe. Most of the mass of the solar system is concentrated in the sun – more than 0.9998 of the total mass of the solar system. One can treat the planets, the masses of which are negligible as compared to the sun, as mass points moving in the static gravitational field of the sun. Inside the solar system one can neglect the static gravitational potential of the planets and deal only with the gravitational potential of the sun.

Consider a planet, a point with negligible mass, which moves in the static gravitational field of a body of spherical symmetry, in a great distance from this central mass. In a very great distance from this central mass the gravitational field is so weak until it is not felt and we arrive back at the Minkowski metric. These are the conditions that Einstein imposed on the gravitational field of the sun.

The point moves on a geodesic line under the influence of the gravitational field of the sun, which is determined by the components of the metric tensor. The details of Einstein's scheme are discussed in the chapters below.

Einstein calculated the equations of the geodesic lines in this space and compared them with the Newtonian equations of the orbits of the planets in the solar system. He thus checked whether there is correspondence between general relativity and the Newtonian theory.

In Newtonian theory the gravitational attraction is a central force, and all planets move in a constant plane around the sun. Hence in polar coordinates the motion of this plane is dependent on the distance r of the planet from the center, and $\phi$ the angle between the line that connects the planet to the center and a line that is chosen arbitrarily.

One can obtain the orbit equation, and one obtains r as a function of $\phi$ (the distance of the planet from the sun at any given angle). The solution of the Newtonian orbit equation is the equation of an ellipse – an orbit in the plane, and the eccentricity e determines the characteristic of the elliptic orbit.

The perihelion of the orbit is the point in which the planet is closest to the sun. This point is found on the major axis of the ellipse, its longest diameter, the line that runs through the centre and both its foci. This major axis was found to slowly turn around the sun; and the Perihelion turned as well. This is the precession of the perihelion, and it is more pronounced the more the eccentricity e is larger.

In Einstein's theory the geodesic equation leads to an orbit equation. The geodesic equation led Einstein to a relativistic equation of the orbit [equation (11)].[37] Einstein found that the difference between the Newtonian orbit equation and the relativistic orbit equation was in an additional term: $2GM/c^2r^3$ that appears in the relativistic equation.

He thus treated first the Newtonian solution to this equation as a first approximation. He then checked, what was the size of the correction that resulted from the addition of this term? He integrated the Newtonian orbit equation first.

The Newtonian solution to the Newtonian orbit equation describes the following: an ellipse of a planet, for which the direction of the major axis and the perihelion should both stay fixed.

Einstein next added the perturbation of the additional term $2GM/c^2r^3$ to this solution in order to see whether the turning of the perihelion resulted from this additional term in the relativistic equation. If this was indeed the result, then the precession of the perihelion would turn to be a result of a relativistic effect, and this was the first triumph of Einstein's new theory.

Einstein obtained a solution which is an ellipse, but this ellipse has a major axis which is not constant and it turns around – and this causes a precession of the perihelion. Einstein wrote the solution to the relativistic equation of the orbit, equation (13) of his November 18 paper, an advance of the perihelion of 43" per century.[38]

On December 22, 1915 Karl Schwarzschild, the director of the Astrophysical Observatory in Potsdam, wrote Einstein from the Russian front.[39] Already on June 1913 Schwarzschild set an eye on Einstein and he persuaded people in the Prussian Academy to pay Einstein the full 12,000 marks. [40] Schwarzschild set out to rework Einstein's calculation in his paper of November 18 of the Mercury perihelion problem. He first responded to Einstein's solution for the first order approximation (4b), and found another first-order approximate solution. The problem would be then physically undetermined, said Schwarzschild, if there are a few approximate solutions.

Subsequently, Schwarzschild went over to a complete solution. He said he realized that there was only one line element, which satisfied the conditions that Einstein imposed on the gravitational field of the sun, as well as Einstein's field equations and determinant condition from the November 18 paper. [41] The problem with Schwarzschild's line element was that a mathematical singularity was seen to occur at the origin. If we consider Schwarzschild's line element, then one easily arrives at Einstein's relativistic equation of the orbit (11); and this equation gives the observed precession of the perihelion of Mercury.

Did Einstein arrive at some form of the exact Schwarzschild solution? Writing Einstein's approximate metric solution, equation (4b) from his November 18 paper in the form of a line element, and then writing it in spherical coordinates, leads to a first order approximation to the Schwarzschild exact solution.[42] Stachel presumes Einstein very likely had not arrived at such a solution before Schwarzschild.[43]

It is interesting to note that Einstein did not include the Schwarzschild exact solution in his 1916 review article, which was written *after* Schwarzschild had found his complete exact solution to Einstein's field equations; a solution which satisfied the same conditions as Einstein's first order approximate solution. Einstein preferred in his 1916 paper to write his November 18 first order approximate solution. Why would Einstein do this? This approximate solution led him to conclude "Thus Euclidean geometry does not apply even to a first approximation in the gravitational field, if we wish to take one and the same rod, independently of its location and orientation, as a realization of the same interval".[44] Einstein needed the first approximation solution to arrive at this conclusion, as discussed later in the context of the 1916 review paper.

Back to the week between November 18 and November 25, 1915; after or while working on the solution of the problem of the Perihelion of Mercury, Einstein could resolve the final difficulties in his November 11 theory. It took him an extra week to arrive at the November 25 field equations. On November 26 Einstein wrote his close friend Zangger, however, only *one* colleague has really understood it [his theory], and he is seeking to clearly "nostrify" it (Abraham's expression).[45] This colleague was David Hilbert.

Recall that on November 19 Hilbert sent Einstein a letter in which he congratulated him on overcoming the perihelion motion. Hilbert ended his letter by asking Einstein to continue and keep him up to date on his latest advances. Hilbert did not tell Einstein about the important talk he was giving the day afterwards. Hilbert presented on November 20 a paper to the Göttingen Academy of Sciences, "The Foundations of Physics", including his version to the gravitational field equations of general relativity. Five days later on November 25, Einstein presented to the Prussian Academy his version to the gravitational field equations. An analysis of the priority dispute is brought in detail here.

The conclusions of this study are that at the end of the day it appears that Einstein did not "nostrify" Hilbert. After November 18 Einstein was no more influenced by Hilbert's Weltanschauung and theory of matter, and he was thus not in competition with him anymore.[46] His new field equations of November 25 with the new trace term *are related to his work of November 4, and appear to have sprung from it,* as shown in in this study after discussing Einstein's November 4, 11 and 18 Arbeits.

And what about Einstein's feeling that he expressed to Zangger, according to which Hilbert was seeking to "nostrify" his theory (the November 4 Arbeit that he had sent

Hilbert)? In the printed December version of Hilbert's November 20 paper, Hilbert (willingly?...) acknowledged Einstein's priority, and thus it was after all as usual with Einstein, a help he had finally (willingly?) given to his colleagues (while competing with them).

## 2 First Talk, November 4, 1915: "On the General Theory of Relativity"

On November 4, 1915 Einstein wrote his elder son Hans Albert Einstein, "In the last days I completed one of the finest papers of my life; when you are older I'll tell you about it".[47] The day this letter was written Einstein presented this paper to the Prussian Academy of Sciences. The paper, "Zur allgemeinen Relativitätstheorie", was the first out of four papers that corrected his "Entwurf" 1914 review paper.[48] Einstein's work on this paper was so intense during October 1915 that he told Hans Albert in the same letter, "I am often so in my work, that I forget lunch".[49]

In the first paper of November, the November 4 paper, Einstein gradually expanded the range of the covariance of his field equations. Every week he expanded the covariance a little further until he arrived on November 25 to fully generally covariant field equations.

### 2.1. The Hamiltonian (78) of 1914

In the introduction to his November 4, 1915 paper Einstein explained to his audience at the Prussian Academy of Sciences that his efforts were basing a general theory of relativity, also for nonuniform motion, upon the supposition of relativity. Einstein believed he had found the only law of gravitation that complies with a reasonably formulated postulate of general relativity. He tried to demonstrate the truth of this solution, in "a paper that appeared last year", that is, in Part D of his "Entwurf" 1914 review paper.[50]

The first problematic element that occupied Einstein straight in the introduction to his paper was the 1914 "Hamiltonian" H, equation (78).[51] In his 1914 review paper Einstein was evidently quite pleased with his accomplishment of providing a proof that the H, the variation of which leads to the "Entwurf" field equations, is uniquely fixed by the requirement that it be invariant under the restricted covariance group; and thus Einstein thought he had demonstrated that his "Entwurf" field equations were the only equations that were invariant under the restricted covariance group.[52]

However, his triumph did not last very long. [53] He wrote to Lorentz three weeks before presenting the November 4 paper to the Prussian Academy that, only the choice of (78) enabled the 1914 theory to comply with the Newtonian limit (correspondence principle). However, he discovered that this choice was an error.[54]

Einstein repeated this claim in his introduction to his November 4, 1915 paper.[55] He gave the following reasoning, "The postulate of relativity, *as far as I demanded there*, is always satisfied if the Hamiltonian principle underlies as a basis; but in reality, it provides no tool to establish the Hamiltonian function H of the gravitational field. As a matter of fact, expressing the choice of limiting H, equation (77) a.a.0, says nothing else than that H should be invariant with respect to linear transformations, a demand that has nothing to do with the relativity of acceleration. Furthermore, the choice taken by equation (78) a.a.0, does not determine equation (77)".[56]

Einstein then said in the introduction to the November 4 paper, "For these reasons, I completely lost trust in my established field equations, and looked for a way to limit the possibilities in a natural manner. Thus I arrived back at the demand of a broader general covariance for the field equations, from which I parted, though with a heavy heart, three years ago when I worked together with my friend Grossmann. As a matter of fact, we then have already come quite close to the solution of the problem given in the following".[57]

Einstein concluded the introduction by explaining that, just as the special theory of relativity is based upon the postulate that all equations have to be covariant relative to linear orthogonal transformations, so the theory developed by him in this paper rests upon the postulate of *the covariance of all systems of equations relative to transformations with the substitution determinant 1*.

At the end of the introduction to the November 4 paper, after dropping the Einstein-Grossmann theory and adopting the postulate of determinate 1, Einstein said he grasped the charm and saw the significance of the triumph of the general differential calculus as founded by Gauss, Riemann, Christoffel, Ricci, and Levi-Civita.[58]

It took Einstein three years of bumpy route to grasp the charm of this differential calculus, because he was developing the physics of general relativity, but also using new and yet not fully developed mathematical tools of differential calculus.[59]

### 2.2. The postulate: only substitutions of determinant 1 are permitted

In section §1 "Laws Forming Covariants" Einstein presented the methods of absolute differential calculus, and corrected, and based himself on the detailed presentation and equations from his 1914 "Entwurf" review paper. He was then about to gradually link himself with general covariance theory, and "Riemann's covariant". Einstein could find all the missing ingredients in his 1914 review paper and his Zurich Notebook. He did not need anything else. He said that he could be brief when presenting the laws of forming covariants. He needed only to examine what *will change* in the theory of covariants if only assuming the postulate that substitutions of determinant 1 are permitted.[60]

Section §1 can therefore assist in recovering Einstein's switch from the 1914 "Entwurf" field equations to his November 4 equations.

**2. 3. Einstein's October-November changes**

Einstein referred to equations (14) and (17) [or equation (17a)] of his 1914 paper. [61] This was the problematic starting point.

According to equation (14) Einstein realized that the determinant $|\alpha_{\sigma\mu}|$ could be equated to 1. He thus wrote that the following equation, which is valid for any substitutions,

$$d\tau' = \frac{\partial(x'_1 \ldots x'_4)}{\partial(x_1 \ldots x_4)} d\tau$$

becomes under the premise that only substitutions of determinant 1 are permitted,

$$(1)\quad \frac{\partial(x'_1 \ldots x'_4)}{\partial(x_1 \ldots x_4)} = 1,$$

the equation:

$$(2)\quad d\tau' = d\tau.$$

And the four-dimensional volume element dτ is therefore invariant.

According to equation (16) from the 1914 paper,

$$(16)\ |g_{\mu\nu}| = |\alpha_{\mu\nu}|^2 = 1.$$

Einstein thus concluded that equation (17) [or (17a)] $\sqrt{-g}d\tau$ is an invariant with respect to arbitrary substitutions, and it follows that:

$$(3)\ \sqrt{-g'} = \sqrt{-g}.$$

Thus $|g_{\mu\nu}|$ is therefore an invariant. $\sqrt{-g}$ is then a scalar, and this could lead to a simplification of the basic formulas of the formation of covariants.[62]

In his 1914 paper the factors $\sqrt{-g}$ and $1/\sqrt{-g}$ occurred in Einstein's most basic formulas. Einstein realized that in a theory of covariants, in which only substitutions of determinant 1 are allowed, the factors $\sqrt{-g}$ and $1/\sqrt{-g}$ no longer occur in the basic formulas, and the distinction between tensors and V-tensors is eliminated.[63]

After omitting the factor $\sqrt{-g}$ Einstein showed that equations (19) and (21) of his 1914 paper could be written as one equation:[64]

$$(4)\ G_{iklm} = G^{iklm} = \delta_{iklm}$$

As to equations (29) and (30) of his 1914 paper, Einstein understood he could not simplify them. These were actually originally written by Christoffel. However, Einstein managed to simplify equations (30) and (31). Einstein simplified (31) to an equation (5). He then simplified this equation to a simple definition of divergence (5a).[65]

According to equations (24) and (24a) Einstein wrote instead of (33):

$$(6)\ \Sigma_\tau \begin{Bmatrix} s\tau \\ s \end{Bmatrix} = \frac{1}{2}\Sigma_{\alpha s}\, g^{s\alpha}\left(\frac{\partial g_{s\alpha}}{\partial x_\tau} + \frac{\partial g_{\tau\alpha}}{\partial x_s} - \frac{\partial g_{s\tau}}{\partial x_\alpha}\right) = \frac{1}{2}\Sigma\, g^{s\alpha}\frac{\partial g_{s\alpha}}{\partial x_\tau} = \frac{\partial(\lg\sqrt{-g})}{\partial x_\tau}$$

Einstein also replaced his 1914 equation (37) with a simpler one by omitting the $\sqrt{-g}$.[66] And using the same assumption, in place of equation (41a) Einstein wrote:

$$(9)\ A_\sigma = \sum_\nu \frac{\partial A_\sigma^\nu}{\partial x_\nu} - \frac{1}{2}\sum_{\mu\nu\tau} g^{\tau\mu}\frac{\partial g_{\mu\nu}}{\partial x_\sigma} A_\tau^\nu.$$

A comparison with equation (41b) shows that the law of divergence is the same as that for the divergence of V-tensors in the general differential calculus. Einstein emphasized that this remark applied to any divergence of tensors, as can be derived from (5) and (5a).[67]

Einstein limited himself to transformations of determinant equal to 1, and he realized that this step simplified his 1914 equations. Moreover, he obtained covariants which were formed only from the $g_{\mu\nu}$ and their derivatives.

**2.4. The Ricci Tensor**

Einstein was led to the Riemann-Christoffel tensor of rank four (ik, lm). However, in gravitation Einstein was interested in tensors of rank two, the Ricci tensor $G_{im}$, which could be obtained from the Riemann tensor by contraction, or multiplication with $g_{\mu\nu}$:[68]

$$(12)\ G_{im} = \sum_{kl} g^{kl}(ik, lm).$$

Einstein noted that the derivation of this tensor was better obtained from a different form of the Riemann-Christoffel tensor (ik, lm), [69]

$$\{ik, lm\} = \sum_{\rho} g^{k\rho}(i\rho, lm) = \frac{\partial \begin{Bmatrix} il \\ \kappa \end{Bmatrix}}{\partial x_m} - \frac{\partial \begin{Bmatrix} im \\ \kappa \end{Bmatrix}}{\partial x_l} + \sum_{\rho} \left[ \begin{Bmatrix} il \\ \rho \end{Bmatrix} \begin{Bmatrix} \rho m \\ \kappa \end{Bmatrix} - \begin{Bmatrix} im \\ \rho \end{Bmatrix} \begin{Bmatrix} \rho l \\ \kappa \end{Bmatrix} \right].$$

Einstein may have been alluding to the form found in his Zurich Notebook. [70]

Contracting the Riemann tensor results in the Ricci tensor $G_{im}$, [71]

(13) $G_{im} = \{il, lm\} = R_{im} + S_{im}$.

This division of equation (13) was indeed already implicitly obtained by Einstein in page 22R of his Zurich Notebook.[72] Einstein then equated $R_{im}$ to the second term like he had done three years earlier on page 22R: [73]

$$(13a)\ R_{im} = -\frac{\partial \begin{Bmatrix} im \\ l \end{Bmatrix}}{\partial x_l} + \sum_{\rho} \begin{Bmatrix} il \\ \rho \end{Bmatrix} \begin{Bmatrix} \rho m \\ l \end{Bmatrix}$$

And the first and third terms:

$$(13b)\ S_{im} = \frac{\partial \begin{Bmatrix} il \\ l \end{Bmatrix}}{\partial x_m} - \begin{Bmatrix} im \\ \rho \end{Bmatrix}\begin{Bmatrix} \rho l \\ l \end{Bmatrix}.$$

Under the *constraint* of equation (1), that is, to transformations with determinant equal to 1, $G_{im}$, $R_{im}$ and $S_{im}$ were all tensors.

It follows from the fact that $\sqrt{-g}$ is a scalar, and $\begin{Bmatrix} il \\ l \end{Bmatrix}$ is a covariant four vector,

because of equation (6). Due to (29) from Einstein's 1914 paper, $S_{im}$ is an extension of this four-vector, and thus it is also a tensor. Since $G_{im}$ and $S_{im}$ are tensors, it follows from (13) that $R_{im}$ is also a tensor. Einstein noted that the tensor $R_{im}$ was of utmost importance for gravitation. [74] It actually replaced Einstein's problematic 1914 gravitation tensor $\frac{\mathfrak{G}_{\mu\nu}}{\sqrt{-g}}$.

**2.5. The Components of the Gravitational Field**

In section §2 "Notes on the Differential Laws of 'Material' Processes", Einstein started with the energy-momentum theorem for matter (including electromagnetic processes in the vacuum). [75]

According to the considerations of section §1, there is no distinction between V-tensors and tensors; Einstein therefore rewrote equation (42a) from his 1914 paper as equation with ordinary tensors:

$$(14)\ \sum_{\nu} \frac{\partial T_\sigma^\nu}{\partial x_\nu} = \frac{1}{2} \sum_{\mu\tau\nu} g^{\tau\mu} \frac{\partial g_{\mu\nu}}{\partial x_\sigma} T_\tau^\nu + K_\sigma$$

$T_\sigma^\nu$ is an ordinary tensor and $K_\sigma$ and ordinary four-vector (not a V-tensor and a V-vector). [76]

Einstein then explained, [77]

"This equation of conservation [(14)] led me in the past to view the quantities

$$\frac{1}{2}\sum_{\mu} g^{\tau\mu} \frac{\partial g_{\mu\nu}}{\partial x_{\sigma}}$$

as the natural expression of the components of the gravitational field, even though in view of the formulas of the absolute differential calculus, it is better to introduce the Christoffel symbols

$$\begin{Bmatrix} \nu\sigma \\ \tau \end{Bmatrix}$$

instead of these quantities. This was a fateful prejudice. A preference for the Christoffel symbols is justified especially because of the symmetry in their two covariant indices (here ν and σ) and, because the same thing occurs in the fundamental important equations of the geodesic line (23b) a.a.0, which, seen from the physical point of view, are the equations of motion of a material point in a gravitational field."

Einstein was misled in his search for the gravitational field equations, a search that took him three years after he had adopted the metric tensor as the mathematical representation of gravity; this was because he regarded the derivatives of the metric tensor (46) of the 1914 paper instead of equation (15a) [below].[78]

Einstein explained this to Sommerfeld on November 28, 1915: [79]

"The key to this solution was my realization that not

$$\sum g^{l\alpha} \frac{\partial g_{\alpha i}}{\partial x_m}$$

but the related Christoffel's symbols $\begin{Bmatrix} im \\ l \end{Bmatrix}$ are to be regarded as the natural expression of the gravitational field 'components'. Once one sees this, then the above equation is very simple, there is no need to transform it for the purpose of a general interpretation by computing the symbols".

And he wrote to Lorentz a month later, on January 1, 1916, the following,[80]

"[…] I had already basically possessed the current equations 3 years ago together with Grossman, who had brought my attention to the Riemann tensor. But because I had not recognized the formal importance of the { } terms, I could not obtain a clear overview and prove the conservation laws [arrive at (14) and (15a)]. I was equally unable to see that Newton's theory was contained in it in first-order approximation; I even thought to have seen the opposite. Then I fell into the jungle!"

The first term on the right-hand side of (14), $\frac{1}{2}\sum_{\mu\tau\nu} g^{\tau\mu}\frac{\partial g_{\mu\nu}}{\partial x_\sigma}T_\tau^\nu$, can be written in the form: $\sum_{\nu\tau}\left\{\begin{matrix}\sigma\nu\\ \tau\end{matrix}\right\}T_\tau^\nu$.

And according to (24) and (24a) from Einstein's 1914 paper and (6) from the November 4 paper, Einstein arrived at the components of the gravitational field: [81]

$$(15a)\ \Gamma_{\mu\nu}^{\sigma} = -\left\{\begin{matrix}\mu\nu\\ \sigma\end{matrix}\right\} = -\sum_\alpha g^{\sigma\alpha}\left\{\begin{matrix}\mu\nu\\ \sigma\end{matrix}\right\}$$
$$= -\frac{1}{2}\sum_\alpha g^{\sigma\alpha}\left(\frac{\partial g_{\mu\alpha}}{\partial x_\nu}+\frac{\partial g_{\nu\alpha}}{\partial x_\mu}-\frac{\partial g_{\mu\nu}}{\partial x_\alpha}\right).$$

$K_\sigma$ vanishes when $T_\sigma^\nu$ denotes the energy tensor of all "material" processes and the conservation theorem takes the form:

$$(14a)\ \sum_\alpha \frac{\partial T_\sigma^\alpha}{\partial x_\alpha} = \sum_{\alpha\beta}\Gamma_{\sigma\beta}^{\alpha}\Gamma_{\alpha}^{\beta}.$$

According to (15a) Einstein wrote the equations of the geodesic line, the equations of motion of a material point (23b) from his 1914 paper, in the following form:

$$(15b)\ \frac{d^2x_\tau}{ds^2} = \sum_{\mu\nu}\Gamma_{\mu\nu}^{\tau}\frac{dx_\mu}{ds}\frac{dx_\nu}{ds}.$$

**2.6. The Field Equations**

In section §3 Einstein finally derived the new field equations. He started by writing the general form of the field equations:

(16) $R_{\mu\nu} = -\chi T_{\mu\nu}$.

Einstein wrote, "we already know that these equations are covariant with respect to arbitrary transformations of a determinant equal to 1".[82] Hence, (16) were not yet generally covariant, but only covariant for transformations stratifying equation (1). Saying this Einstein knew he was presenting the first paper and this was only the beginning; he was very likely already working on the second part of the theory, expanding covariance. This is especially true in light of the second part of this paper, equation (21a), as discussed below.

Einstein was now using (13a) and (15a), [(14a)] to write (16) in full form: [83]

$$(16a)\ \sum_{\alpha}\frac{\partial T^{\alpha}_{\mu\nu}}{\partial x_{\alpha}}+\sum_{\alpha\beta}\Gamma^{\alpha}_{\mu\beta}\Gamma^{\beta}_{\nu\alpha}=-\chi T_{\mu\nu}.$$

Equations (16) and (16a) are non-linear. On the left-hand side of the equation $R_{\mu\nu}$ includes the metric tensor and its derivatives; on the right-hand side the sources determine these variables, the stress-energy tensor. The equations are non-linear because of $\Gamma^{\sigma}_{\mu\nu}$, which are defined by (15a).

### 2.7. Writing the Field Equations in Hamiltonian Form

After writing equation (16a) Einstein demonstrated that his field equations (16) could be brought into Hamiltonian form. This demonstration was essential for Einstein because it enabled Einstein to show that his field equations (16) "satisfy the conservation laws".[84] Einstein defined a new tensor $\mathfrak{L}$ (the Hamiltonian H) for the

second term of (16a) multiplied with $g^{\sigma\tau}$:[85]

$$\mathfrak{L}=\sum_{\sigma\tau\alpha\beta}g^{\sigma\tau}\Gamma^{\alpha}_{\sigma\beta}\Gamma^{\beta}_{\tau\alpha}.$$

According to (14a) this tensor had to be connected with the conservation theorem.

And the action [which came instead of equation (61a) of the 1914 paper,

$\delta\{\int H\sqrt{-g}d\tau\}$ [86]]:

$$(17)\ \delta\{\int(\mathfrak{L}-\chi\sum_{\mu\nu}g^{\mu\nu}T_{\mu\nu})d\tau\}.$$

The $g^{\mu\nu}$ have to be varied and the $T_{\mu\nu}$ are to be treated as constants.

Thus (17) was equivalent to:[87]

$$(18)\ \sum_{\alpha}\frac{\partial}{\partial x_\alpha}\left(\frac{\partial\mathfrak{L}}{\partial g^{\mu\nu}_{\alpha}}\right)-\frac{\partial\mathfrak{L}}{\partial g^{\mu\nu}}=-\chi T_{\mu\nu}.$$

where, $\mathfrak{L}$ is a function of the $g^{\mu\nu}$ and of the $g^{\mu\nu}_{\alpha}=\frac{\partial g^{\mu\nu}}{\partial x_\alpha}$.

The second term on the left-hand side $\frac{\partial\mathfrak{L}}{\partial g^{\mu\nu}}$ is analogous to (14a):

$$(19)\ \frac{\partial\mathfrak{L}}{\partial g^{\mu\nu}}=\sum_{\alpha\beta}\Gamma^{\alpha}_{\sigma\beta}\Gamma^{\beta}_{\nu\alpha}$$

And the term in the brackets in the first term on the left-hand side $\frac{\partial\mathfrak{L}}{\partial g^{\mu\nu}_{\alpha}}$ is:

$$(19a)\ \frac{\partial\mathfrak{L}}{\partial g^{\mu\nu}_{\alpha}}=\Gamma^{\alpha}_{\mu\nu}.$$

Inserting (19) and (19a) into (18) gives (16a). This step was important because it means that (18) is equivalent to the field equations (16a). Hence Einstein could use the Hamiltonian form (18) to demonstrate that the principle of the conservation of energy and momentum was satisfied by (16a).

Einstein multiplied (18) by $g^{\mu\nu}_{\sigma}$ with summation over the indices μ and ν:[88]

$$[19b]\ \sum_{\alpha\mu\nu}\frac{\partial}{\partial x_\alpha}\left(g^{\mu\nu}_{\sigma}\frac{\partial\mathfrak{L}}{\partial g^{\mu\nu}_{\alpha}}\right)-\frac{\partial\mathfrak{L}}{\partial x_\alpha}=-\chi\sum_{\mu\nu}T_{\mu\nu}g^{\mu\nu}_{\sigma}.$$

Recall equation (14) the new energy-momentum theorem [replacing (42a) of 1914]. (14) for the total energy tensor of matter reads:

$$[19c]\ \sum_{\lambda}\frac{\partial T_{\sigma}^{\lambda}}{\partial x_{\lambda}}=-\frac{1}{2}\sum_{\mu\nu}\frac{\partial g^{\mu\nu}}{\partial x_{\sigma}}T_{\mu\nu}.$$

Einstein defined the energy tensor of the gravitational field ("which by the way has tensorial character only under linear transformations"):

$$(20a)\ t_{\sigma}^{\lambda}=\frac{1}{2\chi}\left(\mathfrak{L}\delta_{\sigma}^{\lambda}-\sum_{\mu\nu}g_{\sigma}^{\mu\nu}\frac{\partial\mathfrak{L}}{\partial g_{\lambda}^{\mu\nu}}\right).$$

Using (19a) – the components of the gravitational field – this could also be written as:

$$(20b)\ t_{\sigma}^{\lambda}=\frac{1}{2}\delta_{\sigma}^{\lambda}\sum_{\mu\nu\alpha\beta}g^{\mu\nu}\Gamma_{\mu\beta}^{\alpha}\Gamma_{\nu\alpha}^{\beta}-\sum_{\mu\nu\alpha}g^{\mu\nu}\Gamma_{\mu\sigma}^{\alpha}\Gamma_{\nu\alpha}^{\lambda}.$$

Using (20a) and the equations [19b] and [19c] Einstein was able to prove conservation of energy-momentum: [89]

$$(20)\ \sum_{\lambda}\frac{\partial}{\partial x_{\lambda}}\left(T_{\sigma}^{\lambda}+t_{\sigma}^{\lambda}\right)=0.$$

This equation replaced equation (42c) from the 1914 paper.[90]

As to the above field equations in Hamiltonian form; Einstein's equation (18) from November 4 replaced Einstein field equation (81) from his 1914 paper:

$$(81)\sum_{\alpha\beta}\frac{\partial}{\partial x^{\alpha}}\left(\sqrt{-g}\,g^{\alpha\beta}\Gamma_{\sigma\beta}^{\nu}\right)=-\chi(\mathfrak{T}_{\sigma}^{\nu}+t_{\sigma}^{\nu})$$

Recall that in the 1914 paper Einstein used equation (46), the components of the action of the gravitational field on a material point; and in November 4 he replaced it with equation (15a) which led to (14a). In 1914 – by analogy with (46) – Einstein wrote the components of the strength of the gravitational field:[91]

$$(81a)\ \Gamma_{\sigma\beta}^{\nu}=\frac{1}{2}\sum_{\tau}g^{\nu\tau}\frac{\partial g_{\sigma\tau}}{\partial x_{\beta}},$$

(81a) entered into (81) in the 1914 paper. Equation (19) of the November 4 paper [which replaced (81a) of the 1914 paper] was analogous to (14a). Equation (19) entered into equation (18) of the November 4 paper.

In the 1914 paper Einstein used equation (81a) and wrote:

$$(81b)\ t_\sigma^\nu = \frac{\sqrt{-g}}{\chi} \sum_{\mu\rho\tau\tau'} \left( g^{\nu\tau} \Gamma^\rho_{\mu\sigma} \Gamma^\mu_{\rho\tau} - \frac{1}{2} \delta^\nu_\sigma g^{\tau\tau'} \Gamma^\rho_{\mu\tau} \Gamma^\mu_{\rho\tau'} \right)$$

Einstein called $t_\sigma^\nu$ the energy tensor of the gravitational field and emphasized that it have tensorial character only under linear transformations. [92]

In November 4 he replaced this equation with (20a) and wrote the same thing on equation (20a). Equations (15a) and (14a) entered into (16a). And equation (19) entered into equation (20a) – exactly as equation (81a) had entered into equation (81b) in the 1914 paper. Hence, again we see that *Einstein had actually discovered the basics already in the 1914 paper*.

### 2.8. Contraction of the Field Equation

Einstein's next mission was to derive two scalar equations that resulted from his field equations (16a) – these were equations (21) and (22) below; from these he would try to obtain the Newtonian limit. Einstein was going to do that by contracting the Ricci tensor in his field equations. Einstein multiplied (16a) by $g^{\mu\nu}$ and summed over $\mu$ and $\nu$, and using (15a) and (6) he got after some rearranging: [93]

$$(21)\ \sum_{\alpha\beta} \frac{\partial^2 g^{\alpha\beta}}{\partial x_\alpha \partial x_\beta} - \sum_{\sigma\tau\alpha\beta} g^{\sigma\tau} \Gamma^\alpha_{\sigma\beta} \Gamma^\beta_{\tau\alpha} + \sum_{\alpha\beta} \frac{\partial}{\partial x_\alpha} \left( g^{\alpha\beta} \frac{\partial \lg \sqrt{-g}}{\partial x_\beta} \right) = -\chi \sum_\sigma T_\sigma^\sigma .$$

Recall equation (20a), which denotes the energy tensor, and it was written in the form of equation (20b) using (19a) –the components of the gravitational field. The second term on the left-hand side of (21) is equivalent to the second term on the right-hand side of equation (20b). Einstein was thus going to combine between the two equations. He first partly contracted (16a) in order to combine between conservation of energy-momentum (20) and (21). This would lead to a coordinate condition posed on (21). The first two terms on the left-hand side of (21) thus vanished. Einstein was then left only with the third term on the left-hand side which was equal to the right-hand side term.

More specifically: Einstein first multiplied (16a) by $g^{\nu\lambda}$ and summed over ν:[94]

$$\sum_{\alpha\nu}\frac{\partial}{\partial x_\alpha}\left(g^{\nu\lambda}\Gamma^\alpha_{\mu\nu}\right)-\sum_{\alpha\beta\nu}g^{\nu\beta}\Gamma^\alpha_{\nu\mu}\Gamma^\lambda_{\beta\alpha}=-\chi T^\tau_\mu.$$

The second term on the left-hand side is the second term on the right-hand side of (20b), and so Einstein could combine between the two and obtain:

$$\sum_{\alpha\nu}\frac{\partial}{\partial x_\alpha}\left(g^{\nu\lambda}\Gamma^\alpha_{\mu\nu}\right)-\frac{1}{2}\delta^\lambda_\mu\sum_{\mu\nu\alpha\beta}g^{\mu\nu}\Gamma^\alpha_{\mu\beta}\Gamma^\beta_{\nu\alpha}=-\chi\left(T^\tau_\mu+t^\lambda_\mu\right).$$

According to (20) the divergence of the right-hand side equals zero, and thus the divergence of the two right-hand side terms of this equation vanish: [95]

$$(22)\quad \frac{\partial}{\partial x_\mu}\left[\sum_{\alpha\beta}\frac{\partial^2 g^{\alpha\beta}}{\partial x_\alpha\,\partial x_\beta}-\sum_{\sigma\tau\alpha\beta}g^{\sigma\tau}\Gamma^\alpha_{\sigma\beta}\Gamma^\beta_{\tau\alpha}\right]=0$$

This equation is the correction of equation (65a) – written further above – from the 1914 paper. Einstein restricted by the coordinate condition (65a) the coordinate systems (to Angepaßte Koordinatensysteme). [96] In the scheme in the November 4 paper Einstein restricted the field equations (16a) by the coordinate condition (22). Thus (22) led to the vanishing of the first two terms of (21): [97]

$$(22a)\sum_{\alpha\beta}\frac{\partial^2 g^{\alpha\beta}}{\partial x_\alpha\,\partial x_\beta}-\sum_{\sigma\tau\alpha\beta}g^{\sigma\tau}\Gamma^\alpha_{\sigma\beta}\Gamma^\beta_{\tau\alpha}=0.$$

And this equation "tells us that the coordinate system must be adapted to the manifold".[98]

Einstein showed that taking into consideration equation (22a) then equation (21) becomes:

$$(21a)\sum_{\alpha\beta}\frac{\partial}{\partial x_\alpha}\left(g^{\alpha\beta}\frac{\partial \lg\sqrt{-g}}{\partial x_\beta}\right)=-\chi\sum_{\sigma}T_\sigma^\sigma.$$

According to (21a) it is impossible to choose a coordinate system in which $\sqrt{-g}=1$. If we choose such a coordinate system, then log 1 = 0 and the scalar of the energy tensor $\sum_\sigma T_\sigma^\sigma$ is set to zero. [99]

**2.9. The Newtonian Approximation**

Einstein considered equation (22a), which showed that the coordinate system must be adapted to the manifold. But according to (21a) it was impossible to choose such a coordinate system in which $\sqrt{-g}=1$ unless $\sum_\sigma T_\sigma^\sigma = 0$. Einstein was going to deal with this problem in the addendum to the November 4 paper.

Einstein ended his November 4 paper with section §4: The first approximation of equation (22a) is the following: [100]

$$\sum_{\alpha\beta}\frac{\partial^2 g^{\alpha\beta}}{\partial x_\alpha\,\partial x_\beta}=0.$$

Einstein then wrote:

"This coordinate system is not yet defined, adding to this would be the determination of the same relevant 4 equations. We can therefore set as a first approximation arbitrarily

$$(22)\ \sum_\beta\frac{\partial g^{\alpha\beta}}{\partial x_\beta}=0".$$ [101]

Einstein said "arbitrarily" (willkürlich), because he used this approximation (the Hertz condition) already in his Zurich Notebook on page 19R:

$$\sum_{\kappa}\gamma_{\kappa\kappa}\frac{\partial g_{i\kappa}}{\partial x_\kappa}=0$$

At the bottom half of page 22R Einstein arrived at a candidate for the left-hand side of the field equations that was extracted from the November tensor by imposing the Hertz condition. [102] On page 23L he abandoned the Hertz condition:

"$\sum \frac{\partial \gamma_{\kappa\alpha}}{\partial x_{\kappa}} = 0$ sei = 0 ist nicht nötig." (not necessary).[103]

But there was still a problem with the Hertz condition as Einstein himself explained it later in a note that he has appended at the foot of the letter to Karl Schwarzschild, "The choice of coordinate system demanded by the condition $\sum \frac{\partial g^{\mu\nu}}{\partial x_{\nu}} = 0$ [the Hertz condition] is not consistent with $\sqrt{-g} = 1$".[104]

In the end of the November 4 paper Einstein used (22) above [and equation (15a)] to rewrite (16a) in the form:

$$\text{(16b)} \quad \frac{1}{2}\sum_{\alpha} \frac{\partial^2 g_{\mu\nu}}{\partial x_{\alpha}^2} = \chi T_{\mu\nu},$$

which contains Newton's law of gravitation (Poisson's equation) as an approximation.[105] Newtonian equations are obtained from the relation between the gravitational force and the potential. Einstein demanded that the general form of his field equations would be similar to Poisson's equation.

**2.10. Einstein's First Letter to Hilbert**

Meanwhile on November 7 Einstein wrote David Hilbert and he sent him the proofs of this paper of November 4: "I am sending you the correction to a paper in which I changed the gravitation equations". Einstein was telling Hilbert that he corrected his 1914 paper into the November 4, 1915 paper that he was sending him, "after I myself realized about 4 weeks ago that my previous method of proof was fallacious". Since Hilbert had found a mistake in Einstein's 1914 paper, Einstein wanted him to look at his new work. [106]

Einstein explained the mistake ("found a hair in my soup, which spoiled it entirely for you"[107]) that Hilbert had found in his 1914 paper in a letter he sent to Hilbert on March 30, 1916.[108] Einstein wrote, "The error you found in my paper of 1914 has now become completely clear to me".[109] In fact Hilbert was troubled by Einstein's theorem and its proof from section §14 of his 1914 paper. Recall that this theorem and proof also occupied Tullio Levi-Civita in his correspondence with Einstein between March 1915 and May 1915. These supplied the formal basis for Einstein's belief that if the coordinate system was an adapted coordinate system, then the gravitational tensor $\mathfrak{G}_{\mu\nu}$ was a covariant tensor.

In the 1914 paper Einstein wrote in his proof the following: [110]

He first wrote equation (61). After varying infinitesimally the $g^{\mu\nu}$ he rewrote this equation in the following form:

$$dJ = \delta\left\{\int H\sqrt{-g}\,d\tau\right\} = \int d\tau \sum_{\mu\nu\sigma}\left\{\frac{\partial(H\sqrt{-g})}{\partial g^{\mu\nu}}\delta g^{\mu\nu} + \frac{\partial(H\sqrt{-g})}{\partial g_{\sigma}^{\mu\nu}}\delta g_{\sigma}^{\mu\nu}\right\},$$

**and since:** $\boldsymbol{\delta g_{\sigma}^{\mu\nu} = \frac{\partial}{\partial x_{\sigma}}(\delta g^{\mu\nu})}$

after partial integration and considering the vanishing of $\delta g^{\mu\nu}$ at the boundary of $\Sigma$,

$$(71)\ \delta J = \int d\tau \sum_{\mu\nu} \delta g^{\mu\nu}\left\{\frac{\partial H\sqrt{-g}}{\partial g^{\mu\nu}} - \sum_{\sigma}\frac{\partial}{\partial x_{\sigma}}\left(\frac{\partial H\sqrt{-g}}{\partial g_{\sigma}^{\mu\nu}}\right)\right\}.$$

Via his theorem Einstein proved that under limitation to adapted coordinate systems $\delta J$ was invariant. Using his theorem thus proved, Einstein concluded that the quantity in the brackets of (71) in the above integral times $\delta g^{\mu\nu}$, and in (72) below [tensor $\mathfrak{G}_{\mu\nu}$] divided by $\sqrt{-g}$ is also an invariant:

$$(72)\ \frac{1}{\sqrt{-g}}\sum \delta g^{\mu\nu}\mathfrak{G}_{\mu\nu}$$

And Einstein tried to save the covariance of his gravitational tensor $\mathfrak{G}_{\mu\nu}$ (under limitation to adapted coordinate systems)[111]

Hilbert showed Einstein that $\delta g_{\sigma}^{\mu\nu} = \frac{\partial}{\partial x_{\sigma}}(\delta g^{\mu\nu})$ is not valid. [112] Hence, from the mathematical point of view, equation (71) is also not valid and so is equation (72) and Einstein's gravitational tensor $\mathfrak{G}_{\mu\nu}$. Levi-Civita also explained to Einstein (while supplying other reasons) that (72) was not valid.

**3. Second Talk, November 11, 1915: "On the General Theory of Relativity (Addendum)"**

Next week, on Thursday, the mathematical-physical class of the Prussian academy gathered again to hear a correction to the November 4 paper. Einstein presented this correction in his Addendum, "Zur allgemeinen Relativitätstheorie (Nachtrag)".

**3.1 New Condition, the Scalar of the Energy Tensor Vanishes**

In the previous paper Einstein obtained equations (16) which are covariant with respect to transformations of determinant equal to 1. Einstein then obtained equation (21a) and the impossibility to choose a coordinate system in which $\sqrt{-g} = 1$, for then $\sum_\sigma T_\sigma^\sigma = 0$. Einstein thus set to solve this problem by dropping the November 4 postulate of determinant 1 and adopting it as a coordinate condition.

The (stress)-energy tensor of "matter" $T_\mu^\lambda$ has a scalar (trace) $\sum_\mu T_\mu^\mu$. It is well known that it vanishes in an electromagnetic field. But it differs from zero for matter proper. Einstein considered as the "simplest special case" an "incoherent" continues (incompressible) fluid. For this case, said Einstein, the trace of the stress-energy tensor does not vanish. [113]

Suppose we reduce matter to electromagnetic processes, and also assume that gravitational fields could be related to "matter"; that is, gravitational fields form an important constituent of "matter". In such a theory the scalar of the energy tensor would also vanish.

In that case, $\sum_\mu T_\mu^\mu$ *appears* to be positive for the entire structure, but it would still vanish. Einstein used instead of $\sum_\mu T_\mu^\mu$ as the stress-energy tensor to represent "matter", $\sum\left(T_\mu^\mu + t_\mu^\mu\right)$ as the stress-energy-tensor to represent "matter" (i.e., the stress-energy tensor is composed of two contributions, electromagnetic + gravitational fields). This combination of the stress-energy tensor enabled Einstein to include $\sum t_\mu^\mu$ that could be positive, while in general the whole expression vanished, and $\sum_\mu T_\mu^\mu$ actually vanishes everywhere. Einstein thus added the coordinate conditions:

"*we assume in the following that the condition* $\sum T_\mu^\mu = 0$ *is in general actually fulfilled* ".[114]

This hypothesis allowed Einstein to take the last step and to write the field equations of gravitation in a general covariant form.

Einstein arrived at the above hypothesis by making new assumptions on electromagnetic and gravitational "matter". What impelled Einstein's change of perspective in the November 11 paper?

Renn and Stachel say the answer seems to lie in Einstein's interaction with Hilbert. It would have been quite uncharacteristic of Einstein to adopt the new approach so readily had it not been for current discussions of the electrodynamic worldview and his feeling that he was now in competition with Hilbert.[115]

Renn and Stachel explain that when one examines Einstein's previous writings on gravitation, published and unpublished, one finds no trace of an attempt to unify gravitation and electromagnetism. He had never advocated the electromagnetic worldview. On the contrary, he was apparently disinterested in Mie's attempt at a unification of gravitation and electrodynamics, not finding it worth mentioning in his 1913 review of contemporary gravitation theories. And soon after completion of the final version of general relativity, Einstein reverted to his earlier view that general relativity could make no assertions about the structure of matter. Einstein's mid-November 1915 pursuit of a relation between gravitation and electromagnetism was, then, merely a short-lived episode in his search for a relativistic theory of gravitation.

Renn and Stachel examined Einstein's correspondence with Hilbert and say it is quite clear that this perspective of a theory of matter was shaped by Einstein's rivalry with Hilbert. It thus seems quite clear that Einstein's temporary adherence to an electromagnetic theory of matter was triggered by Hilbert's work, which Einstein attempted to use in order to solve a problem that had arisen in his own theory, and that he dropped it when he solved this problem in a different way.

Ironically then, Hilbert's most important contribution to general relativity may have been enhancing the credibility of a speculative and ultimately untenable physical hypothesis that guided Einstein's final mathematical steps towards the completion of his theory. [116]

**3.2 The First Generally Covariant Field Equations**

In the previous paper Einstein wrote the following equation, [117]

(13) $G_{im} = \sum_l \{il, lm\} = R_{im} + S_{im}$

and showed that $G_{im}$ is a covariant tensor.

Einstein also wrote in the previous paper: [118]

$$(13a)\ R_{im} = -\sum_{l} \frac{\partial \begin{Bmatrix} im \\ l \end{Bmatrix}}{\partial x_l} + \sum_{\rho l} \begin{Bmatrix} il \\ \rho \end{Bmatrix} \begin{Bmatrix} \rho m \\ l \end{Bmatrix},$$

$$(13b)\ S_{im} = \sum_{l} \frac{\partial \begin{Bmatrix} il \\ l \end{Bmatrix}}{\partial x_m} - \sum_{\rho l} \begin{Bmatrix} im \\ \rho \end{Bmatrix} \begin{Bmatrix} \rho l \\ l \end{Bmatrix}.$$

And he wrote the following equation,[119]

$$(6) \sum_{i} \begin{Bmatrix} il \\ l \end{Bmatrix} = \frac{1}{2} \sum g^{l\alpha} \frac{\partial g_{l\alpha}}{\partial x_i} = \frac{\partial (\lg \sqrt{-g})}{\partial x_i}.$$

Einstein understood after November 4 that $G_{im}$ from (13) could be "the only tensor available for the establishment of generally covariant equations of gravitation.

Setting we realize that the field equations of gravitation should be

$(16b)_{\text{November 11}}$ $G_{\mu\nu} = -\chi T_{\mu\nu}$. "[..].[120]

where, $\chi = 8\pi G$. Equations $(16b)_{\text{November 11}}$ are generally covariant.

(One can see that when the gravitational potential is obtained in the 44$^{th}$ component of the metric tensor, $g_{44} = 1 + \Phi/c^2$. For low velocities and in the weak fields approximation, the equation between the 44$^{th}$ components of the stress-energy tensor and the gravitation tensor is equivalent to the Newtonian equation: $G_{44} = -T_{44}$).

Einstein introduced a new coordinate system in such a way that with respect to this system, $\sqrt{-g} = 1$ holds everywhere. In this case $S_{im}$ vanishes because of (6). Thus

one returns to the system of field equations of November 4:

(16) $R_{\mu\nu} = -\chi T_{\mu\nu}$,

and these equations are satisfied only by transformations of determinant 1 [equation (1) of the November 4 paper]. [121]

### 3.3 Another Condition, $\sqrt{-g} = 1$.

We are then led back to Equation (21a) of the November 4 paper. Equation (21a) leads on choosing $\sqrt{-g} = 1$ to the result: $\sum T_{\mu}^{\mu} = 0$. And indeed Einstein ended his November 11 paper with equation (21a).

The resolution of the paradox – according to Einstein on November 11, 1915 – is in Einstein's hypothesis of electromagnetic matter and the stress-energy tensor that represented "matter" $\sum(T_{\mu}^{\mu} + t_{\mu}^{\mu})$. Einstein assumed that $t_{\mu}^{\mu}$ is due to gravitational fields, and it was part of "matter", which was essential to equation (13). Einstein wrote, "*The generally covariant field equations (16b), formed by our starting point, do not lead to a contradiction only if the hypothesis explained in the introduction applies*. But then we are at the same time entitled to add to our previous field equation the limiting condition:

(21b) $\sqrt{-g} = 1$ ." [122]

In the November 4 paper, equations (16) or (16a) have led by contraction to equation (21a). The important equations in this process were the total energy tensor of matter:

$$[19c] \sum_{\lambda} \frac{\partial T_{\sigma}^{\lambda}}{\partial x_{\lambda}} = -\frac{1}{2} \sum_{\mu\nu} \frac{\partial g^{\mu\nu}}{\partial x_{\sigma}} T_{\tau}^{\nu}$$

and conservation of energy-momentum: [123]

$$(20) \sum_{\lambda} \frac{\partial}{\partial x_{\lambda}} (T_{\sigma}^{\lambda} + t_{\sigma}^{\lambda}) = 0.$$

Equation (21a) from November 4 signified to Einstein that equations (16a) from November 4 could not be general, and these should have been expanded to generally covariant field equations arising from (13).

### 3.4 Einstein's Second Letter to Hilbert

Einstein waited one day, and on Freitag, very likely November the 12th, 1915, he wrote his new friend and colleague David Hilbert from Göttingen. He first thanked him for his kind letter. [124] Hilbert's letter was not survived, but it was a reply to Einstein's letter from November 7, to which Einstein attached the proofs to his November 4 paper. [125]

What could Hilbert tell Einstein in this return letter that he had sent him sometime between November 8 and 10? Could Hilbert tell Einstein that equation (21a) seemed to point to a reconsideration of his November 4 paper? Of course we are unable to answer this question; but in light of Einstein's letter to Hilbert from Friday November 12, this scenario is unreasonable.

Einstein wrote Hilbert on November 12 "the problem has meanwhile made new progress. Namely, it is possible to exact general covariance from the postulate $\sqrt{-g} = 1$; Riemann's tensor then deliver's the gravitation equations directly". [126]

Einstein thus told Hilbert on a progress, and a main finding related to $\sqrt{-g} = 1$,

which he very likely arrived as a result of reconsidering equation (21a). But it is certainly reasonable that Einstein was influenced by Hilbert's possible letter (sent before November 10) when arriving at the above solution, as already discussed above: Einstein had noticed that the condition $\sum T_{\mu}^{\mu} = 0$, which follows from setting

$\sqrt{-g} = 1$ in (21a) can be related to an electromagnetic theory of matter.[127]

Let us summarize the situation as to Hilbert's possible knowledge about Einstein's 1914 and November 4 theories:

On November 7, Einstein said (with regards to his "Entwurf" 1914 paper[128]) that Sommerfeld noted Hilbert had "found a hair in my soup, which spoiled it entirely for you"[129]. On November 12 Einstein sent Hilbert two offprints of his 1914 paper. [130] Thus Hilbert *read* Einstein's 1914 paper *before* November 7, 1915. (And he possessed two extra copies by November 13 when replying to Einstein). Hilbert also heard Einstein Göttingen lectures of summer 1915. He also possessed the proofs of Einstein's November 4 paper *before* November 12, because he sent Einstein a reply letter for the latter's November 7 letter.

Einstein's 1914 review article and "Entwurf" theory *initially* inspired Hilbert in his search for generally covariant field equations. Even Einstein's Hole Argument inspired Hilbert towards a comprehensive axiomatic unifying theory of gravitation and matter.[131] Most of the general theory of relativity of November 4, 1915 had originated in Einstein's paper of 1914 and in the Zurich Notebook of 1912.

On November 13, Hilbert replied to Einstein's letter of November 12. Hilbert was fully immersed in Einstein's problem: "Actually, I first wanted to think of a very palpable application for physicists, namely reliable relations between the physical constants, before obliging with my axiomatic solution to your great problem. But since you are so interested, then this coming Tuesday I will develop, approximately after tomorrow (d. 16 of M.), my whole theory in detail". [132]

Hilbert explained to Einstein the main points of his theory, a unification of gravitation and electromagnetism,[133] and told Einstein that he had already discussed his discovery with Sommerfeld. He wanted next to explain it to Einstein on Tuesday. He invited him to come at 3 or 1/2 past 5. "The Math. Soc. Shall meet at 6 o'clock in the auditorium-house. My wife and I would be very pleased if you stayed with us".[134] Hilbert presented the following talk on Tuesday, November 16, to the Mathematical Society of Göttingen, "Grundgleichungen der Physik" ("The Fundamental Equations of Physics").[135]

However, Hilbert added a note at the end of the letter: "As far as I understand your new pap. [paper of November 4], the solution giv. [given] by you is entirely different from mine […]". In addition Hilbert wrote: "Continuation on page I with the invitation to come for Tuesday, 6 o'clock, many greetings H". Thus he sent Einstein another page.[136]

Einstein replied to Hilbert two days later on November 15. He told Hilbert he could not come at the moment to Göttingen, "but should I wait until I can study your system from the printed paper, for I am very tired and in addition plagued with stomach pains. Please send me if possible, a proof copy of your investigation in order to satisfy my impatience".[137]

In response, Hilbert perhaps sent a copy of the lecture he had given on the subject on November 16, or else a copy of a manuscript of the paper he would present five days later on November 20 to the Royal Society in Göttingen.[138]

It is evident that until November 17 Einstein was still in competition with Hilbert and was influenced by his electromagnetic theory of matter. Indeed Einstein told his close friend Michele Besso, "In these last months I had great success in my work. *Generally covariant* gravitation equations. *Perihelion motions explained quantitatively*. The role of gravitation in the structure of matter".[139]

## 4. Third Talk, November 18, 1915: Perihelion Motion of Mercury

### 4.1. The Entanglement of Einstein's Three Papers

Einstein was already less patient after he had received Hilbert's "System". He replied to Hilbert on November 18 telling him that "Your given system agrees – as far as I can see – exactly with what I found in the last few weeks and have presented to the Academy".[140]

Let us come back to on Hilbert's November 13 letter to Einstein, "As far as I understand your new pap. [paper of November 4], the solution giv. [given] by you is entirely different from mine […]". And Einstein wrote Hilbert above that, his given system – as far as he can see – exactly agrees with what he had found.

Corry, Renn, and Stachel say that "it is clear from this exchange that, at this point, at least one of the two – and probably each – is misunderstanding the other's work".[141]

As far as Einstein understood, he claimed that Hilbert's system of equations (which was mentioned by Hilbert to Einstein already in his November 13 letter to the latter) agreed with his own ones presented in the November 4 paper. From Einstein's point of view this was very likely so, because of equations:[142] (13) $G_{im} = \{il, lm\} = R_{im} + S_{im}$, (13a), (13b), and (16) from Einstein's November 4 paper; Einstein had presumably sent Hilbert these in the proofs to that paper on the November 7 letter to the latter. Einstein felt that his work of November 4 was implicitly contained (or "nostrified") in Hilbert's "System".

The day afterwards Hilbert sent a polite letter in which he congratulated Einstein "on overcoming the perihelion motion. If I could calculate as rapidly as you […]".[143] In fact Einstein did not calculate the result that rapidly. A week after November 11, on the coming Thursday, he presented his work on the Perihelion Motion of Mercury; but the basic calculation was done two years earlier with Besso in the Einstein-Besso manuscript. Einstein transferred the basic framework of the calculation from the Einstein-Besso manuscript; but he corrected it according to his November 11 field equations. [144] In addition, Einstein showed that his theory also predicted twice as strong the curvature of light rays due to gravitational fields as was predicted by Einstein due to his "Enwurf" theory.

Although Einstein rapidly calculated the perihelion shift of Mercury on the basis of his previous work with Besso, the success of his calculation was also based on his November 11 theory. The condition $\sqrt{-g} = 1$, implied by the assumption of an electromagnetic origin of matter, was essential for this calculation, which Einstein considered a striking confirmation of his audacious hypothesis on the constitution of matter, definitely favoring this theory over that of November 4. Thus when writing the Perihelion paper Einstein was still influenced by Hilbert's electromagnetic theory of matter.[145]

On the other hand, Einstein was already departing from Hilbert's theory of matter in the November 18 Arbeit. He there wrote, "through which time and space are deprived of the last trace of objective reality".[146] Thus Einstein began to realize that space and time coordinates have no meaning in general relativity.

After November 11, on Thursday, November 18, Einstein presented to the Prussian Academy his paper, "Erklärung der Perihelbewegung des Merkur aus der allgemeinen Relativitätstheorie" (Explanation of the Perihelion Motion of Mercury from the General Theory of Relativity"). He noted right at the beginning of the paper (in a footnote added after the paper was completed): "In a forthcoming communication it will be shown that this hypothesis [the trace of the stress-energy tensor vanishes] is superfluous. It is only essential that such a choice of reference is possible that the determinant $|g_{\mu\nu}|$ takes on the value – 1. The following investigation is independent of this".[147]

Subsequently Einstein expanded his November 11 paper and presented the final version of his field equations, as will be discussed after presenting the November 18 paper. When did Einstein get the idea to expand his November 11 equations? Before he received Hilbert's "System" or afterwards? Did Hilbert's "System" push Einstein to the final solution of November 25? It appears that the answer to this question should be *no*, as will be discussed after examining Einstein's November 25 field equations.

But before discussing this question, one important ingredient to answering it is the essential entanglement among Einstein's three Arbeits – *the 1914 review paper, the November 4 paper and the 1912 work with Grossmann (the Zurich Notebook, p 22R)*. After discussing the paper dealing with the Perihelion Motion of Mercury, it will be shown here that the final solution of the problem, that was presented by Einstein on November 25, was dependent on this entanglement among these three papers, *and not anymore on Hilbert's theory of matter.*[148] Einstein succinctly explained this issue to Hilbert in his letter to the latter *dating from November 18*:

"The difficulty was not in finding generally covariant equations for the $g_{\mu\nu}$'s; for this is easily achieved with the aid of Riemann's tensor. Rather, it was hard to recognize that these equations are a generalization, that is, a simple and natural generalization of Newton's law. I did it only in the last few weeks (my first communication that I sent you [November, 4]), whereas I considered the only possible generally covariant equations, which have now been proven to be correct, 3 years ago with my friend Grossmann [Zurich Notebook]. With heavy heart we separated from them, because it seemed to me that the physical discussion revealed an incompatibility with Newton's law. – The important thing is that the difficulties have now been overcome. I am presenting today to the Academy a paper in which I derive quantitatively out of general relativity, without any guiding hypothesis, the perihelion motion of Mercury discovered by Le Verrier. This was not achieved until now by any gravitational theory".[149]

**4.2. The Gravitational Field of the Sun**

Einstein started his calculations of November 18 with the static gravitational field of the sun (in vacuum), which has to satisfy, upon suitably choosing the reference frame, the following field equations (16b) of the November 11 paper: [150]

$$(1)\quad [(16b)]\sum_{\alpha}\frac{\partial \Gamma^{\sigma}_{\mu\nu}}{\partial x_{\alpha}}+\sum_{\alpha\beta}\Gamma^{\sigma}_{\mu\beta}\Gamma^{\beta}_{\nu\alpha}=0,$$

These are (16a) of the November 4 paper or (16) of the November 11 paper once one introduces a coordinate system in such a way that with respect to this system, $\sqrt{-g}=1$ [(21b)].

And the components of the gravitational field are: [151]

$$(2)\quad [(15)]\ \Gamma^{\sigma}_{\mu\nu}=-\begin{Bmatrix}\mu\nu\\ \alpha\end{Bmatrix}=-\sum_{\beta}g^{\alpha\beta}\begin{Bmatrix}\mu\nu\\ \beta\end{Bmatrix}$$
$$=-\frac{1}{2}\sum_{\beta}g^{\alpha\beta}\left(\frac{\partial g_{\mu\beta}}{\partial x_{\nu}}+\frac{\partial g_{\nu\beta}}{\partial x_{\mu}}-\frac{\partial g_{\mu\nu}}{\partial x_{\alpha}}\right).$$

In light of the field equations (1), Einstein added the hypothesis from his November 11 paper that the trace of the stress-energy tensor of "matter" always vanishes; and he imposed the determinant condition: [152]

$$(3)\quad |g_{\mu\nu}|=-1.$$

Einstein assumed that the sun was a point mass at rest, which was located at the origin of the coordinate system. It produced a gravitational field that could be calculated from the above equations (1) by successive approximations.

Einstein explained that, we should be well aware that the $g_{\mu\nu}$ for a given solar mass are still not completely mathematically determined by equations (1) and (3). This follows from the fact that these equations are covariant with respect to arbitrary transformations with determinant equal to 1. But one can assume that by such successive transformations, however, all these solutions could be reduced to one another, and one could show that they (for given boundary conditions) differ from one another only formally but not physically. "Following this conviction, I shall confine myself, for the time being, here to derive a solution, without entering into the issue of whether it is the only possible one".[153]

In what followed, Einstein intended to obtain a solution, without considering the question whether or not the solution was the only possible unique solution. If Einstein arrived at the result for the advance of the perihelion of Mercury which was close to 45", then his method of using an approximate rather than an exact and unique solution could not be criticized.[154] The first to offer an exact solution to Einstein's field equations was Schwarzschild, as discussed after presenting Einstein's result.

Einstein proceeded to obtain the $g_{\mu\nu}$ for the mass of the sun by the basic method from the 1913 Einstein-Besso manuscript; he transferred this method to his November 18, 1915 paper, and corrected it according to his new November 11, 1915 theory.

**4.3. The 0$^{th}$ approximation of the metric field $g_{\mu\nu}$ of the sun.**

Einstein started from the 0$^{th}$ approximation. $g_{\mu\nu}$ corresponded to the "original" **(special) theory of relativity**, to the flat Minkowski metric: [155]

(4) $g_{\mu\nu} = \text{diag}\,(-1, -1, -1, +1)$,

And Einstein wrote this succinctly: [156]

$$(4a)\ \ g_{\rho\sigma} = \delta_{\rho\sigma},\ g_{\rho 4} = g_{4\rho} = 0,\ g_{44} = 1$$

Here ρ and σ signify indices, 1, 2, 3; the Kronecker delta $\delta_{\rho\sigma}$ is equal to 1 or 0 when $\rho = \sigma$ or $\rho \neq \sigma$, respectively.

The approximation given in equation (4a) forms the 0$^{th}$ approximation.

Einstein then assumed the following: $g_{\mu\nu}$ differ from the values given in equation (4a) by an amount that is small compared to 1. Einstein treated this deviation as a small change of "first order". Functions of n$^{th}$ degree of this deviation were treated as quantities of the n$^{th}$ order. Einstein next used equations (1) and (3) in light of equations (4a) for calculation through successive approximations of the gravitational field[157] of the sun up to quantities of n$^{th}$ order exactly. [158]

In the Einstein-Besso manuscript, Einstein and Besso calculated the metric field $g_{\mu\nu}$ produced by a static sun (spherical mass distribution). Then they calculated the advance of the perihelion in the field of this static sun.

In November 1915 Einstein followed the same reasoning. The metric field $g_{\mu\nu}$ (the solution) had the following four properties, which were four conditions on the gravitational field of the sun: [159]

1) **The solution is static**: all components of the solution are independent of $x_4$ (time coordinate).

2) The solution $g_{\mu\nu}$ is spherically symmetric about the origin of the coordinate system.

3) The equations $(4a)\ g_{\rho 4} = g_{4\rho} = 0$ are valid exactly for ρ = 1, 2, 3.

4) At infinity the $g_{\mu\nu}$ **tends to the values of the Minkowski flat metric of special relativity** given by (4a).

**4.4. The First approximation of the metric field $g_{\mu\nu}$ of the sun.**

To first order, the equations (1) and (3) and the four above conditions are satisfied through the following transformations from:[160]

$$g'_{\rho\sigma} = \mathrm{diag}\left(-\left[1+\frac{\alpha}{r}\right], -1 - 1, 1-\frac{\alpha}{r}\right).$$

to the following solution: [161]

$$(4b) \qquad g_{\rho\sigma} = -\delta_{\rho\sigma} - \alpha\frac{x_\rho x_\sigma}{r^3}, \ \ g_{44} = 1 - \frac{\alpha}{r}.$$

The $g_{\rho\sigma}$ tends to the Minkowski metric (4a) according to condition 4, and the $g_{4\rho}$ and $g_{\rho 4}$ are determined by condition 3.

The $r = \sqrt{x_1{}^2 + x_2{}^2 + x_3{}^2}$, and α (=$2GM/c^2$, in Einstein's paper c = 1) is a constant **determined by the mass of the static sun**.

This is the covariant metric field $g_{44}$ of the sun to first order.

From equation (4b), the theory implies that in the case of a mass at rest, the components $g_{11}$ to $g_{33}$ already differ from zero quantities to first order. Einstein told the reader that "we will see later, however, that no contradiction to Newton’s law (in the first order approximation) arises".[162]

However, Einstein found that his theory laid so far, could lead to another result that occupied him since 1907; it could produce,[163]

"a somewhat different influence of the gravitational field on the light ray as in my earlier work, because the velocity of light is determined by the equation

$$(5)\quad [ds^2 =] \sum g_{\mu\nu}\, dx_\mu dx_\nu = 0.$$

By application of the Huygens principle, we find from equations (5) and (4b) through a simple calculation, that a light ray passing at a distance Δ undergoes an angular deflection of magnitude 2α/Δ, while the earlier calculation, which was not based upon the hypothesis $\sum T_\mu^\mu = 0$, had given the value α/Δ. A light ray grazing the surface of the sun should experience a deflection of 1.7 sec of arc instead of 0.85 sec of arc".

Einstein has calculated the $g_{\mu\nu}$ in the first order approximation. He was now ready to calculate the components $\Gamma^\tau_{\rho\sigma}$ of the gravitational field of the sun to the first order approximation. The first order solutions (4b) are substituted in the equations (2) of the components of the gravitational field of the sun: [164]

$$(6a)\Gamma^\tau_{\rho\sigma} = -\alpha\left(\delta_{\rho\sigma}\frac{x_\tau}{r^2} - \frac{3}{2}\frac{x_\rho x_\sigma x_\tau}{r^5}\right),$$

where, ρ, σ, τ are the indices and are equal to the values 1, 2, 3 and for the 44 component:

$$\textbf{(6b)}\ \boldsymbol{\Gamma^\sigma_{44} = \Gamma^4_{4\sigma} = -\frac{\alpha}{2}\frac{x_\sigma}{r^3}},$$

where, σ is equal to the values 1, 2, 3. The components in which the index 4 appears once or three times vanish. Thus,

$$-\frac{\alpha}{2}\frac{x_\sigma}{r^3}\left(2r\delta_{\rho\sigma} - 3\frac{x_\rho x_\tau}{r^2}\right).$$

**4.5. The Second Order approximation of the metric field $g_{\mu\nu}$ of the sun.**

Einstein needed to determine only three components (σ = 1, 2, 3) $\Gamma^\sigma_{44}$ accurately to quantities of second order to be able to determine accurately the orbits of the planets and the advance of the Perihelion in the gravitational field of the sun. Einstein needed for this the last field equation that he obtained (6b), and the four conditions leading to his solution.

Einstein wrote the second order field equations (1) for the components $\Gamma^\sigma_{44}$: [165]

$$\sum_{\sigma}\frac{\partial\Gamma_{44}^{\sigma}}{\partial x_{\sigma}}+\sum_{\sigma\tau}\Gamma_{4\tau}^{\sigma}\Gamma_{4\sigma}^{\tau}=0.$$

Using equation (6b), and neglecting quantities of third and higher orders, this equation reduces to:

$$\sum_{\sigma}\frac{\partial\Gamma_{44}^{\sigma}}{\partial x_{\sigma}}=\frac{\alpha^2}{2r^4}$$

From this, and using (6b), Einstein obtained the value for the components of the gravitational field of the static sun to the second order approximation

$$\left(-\frac{\alpha}{2}\frac{x_\sigma}{r^3}+\frac{\alpha^2 x_\sigma}{2r^4}\right):$$

$$\textbf{(6c)}\quad \boldsymbol{\Gamma_{44}^{\sigma}=-\frac{\alpha}{2}\frac{x_\sigma}{r^3}\left(1-\frac{\alpha}{r}\right)}.$$

### 4.6. The Motion of the Planets in the Gravitational Field of the Sun

The next step was writing equations of motion for a point mass moving in the gravitational field of the sun to second order. To do so Einstein needed the geodesic equation (15b) from his November 4 paper: a planet in a free fall in the gravitational field of the sun moves on a geodesic line according to the geodesic equation (15b): [166]

$$(7)\ \ [(15)]\ \ \frac{d^2x_\nu}{ds^2}=\sum_{\sigma\tau}\Gamma_{\sigma\tau}^{\nu}\frac{dx_\sigma}{ds}\frac{dx_{\nu\tau}}{ds}.$$

Einstein first said that from this equation we conclude that the Newtonian equation of motion is contained in it as a first approximation. He explained this further: when the speed of the planet is small with respect to the speed of light, then $dx_1$, $dx_2$, $dx_3$ are small as compared to $dx_4$. It follows from this that we get back the first order approximation (4b), in which we always take on the right-hand side the condition $\sigma=\tau=4$.

Then using equation (6b) we get: [167]

$$\boldsymbol{(7a)\ \frac{d^2x_\nu}{ds^2}=\Gamma_{44}^{\nu}=-\frac{\alpha}{2}\frac{x_\nu}{r^3}\ (\nu=1,2,3),\ \frac{d^2x_4}{ds^2}=0}.$$

To first approximation one can set s = $x_4$. Then the first three equations are exactly the Newtonian equations.

Once Einstein recovered the above Newtonian form, he followed the Newtonian procedure, and introduced polar coordinates r and ϕ. Einstein wrote for the orbit of the planet the Newtonian law of conservation of energy and the area law, respectively:[168]

$$(8)\quad \frac{1}{2}u^2 + \Phi = A, \qquad (10)\qquad r^2\frac{d\phi}{ds} = B,$$

where A and B are constants of the energy, and the Newtonian potential is,[169]

$$(8a)\ \Phi = -\frac{\alpha}{2r}, \qquad u^2 = \frac{dr^2 + r^2 d\phi^2}{ds^2}.$$

Recall that $\alpha = 2GM/c^2$ (here $c^2 = 1$). And according to (8a), $\Phi = -GM/r$.

Thus according to equation (8a) equation (8) can be written in the following form:

$$(8)\quad \frac{1}{2}u^2 - \frac{\alpha}{2r} = A$$

Einstein then could evaluate the equations to the next order of approximation. The last of equations (7) yield together with equations (6b),

$$\frac{d^2x_4}{ds^2} = 2\sum_{\sigma}\Gamma^4_{\sigma 4}\frac{dx_\sigma}{ds}\frac{dx_4}{ds} = -\frac{dg_{44}}{ds}\frac{dx_4}{ds},$$

and to first order they lead using (4b) (the value for $g_{44}$) exactly to,[170]

$$(9)\quad \frac{dx_4}{ds} = 1 + \frac{\alpha}{r}.$$

Einstein obtained in a few steps using (7), (6c), (9) and (8) for the equations of motion to the second order the exact following form:

$$(7b)\quad \frac{d^2x_\nu}{ds^2} = -\frac{\alpha}{2}\frac{x_\nu}{r^3}\left(1 + \frac{\alpha}{r} + 2u^2 - 3\left(\frac{dr}{ds}\right)^2\right),$$

where, $u^2 = \delta_{\sigma\tau}\frac{dx_\sigma}{ds}\frac{dx_\tau}{ds}$, and $\left(\frac{dr}{ds}\right)^2 = \frac{dx_\sigma}{ds}\frac{dx_\tau}{ds}$.

This together with equation (9) determines the motion of the mass point in the gravitational field.

Einstein remarked that equations (7b) and (9) give no deviation for the case of an orbital motion from Kepler's third law. Einstein added that from (7b) follows the exact validity of equation (10): "The area law holds in the second order exactly, if the 'proper time' of the planet is used for measuring time". [171]

### 4.7. The Perihelion Advance of Mercury

To obtain the secular rotation of the orbital ellipse from equation (7b), Einstein inserted the terms of first order in the parentheses of the right-hand side of (7b), and he used equation (10) and equation (8) and (8a) for this. By this procedure the terms of the second order on the right-hand side of equation (7b) were not changed.

By this procedure the parentheses of equation (7b) become: [172]

$$\left(1 - 2A + \frac{3B^2}{r^2}\right).$$

Or:

$$\frac{d^2 x_\nu}{ds^2} = -\frac{\alpha}{2}\frac{x_\nu}{r^3}\left(1 - 2A + \frac{3B^2}{r^2}\right).$$

Einstein defined the time variable as $s = \left(s\sqrt{1-2A}\right)$. If B has now a slightly different meaning, then one arrives at an equation which has the form of the Newtonian equations of motion: [173]

$$(7c)\ \frac{d^2 x_\nu}{ds^2} = -\frac{\partial \Phi}{\partial x_\nu}, \qquad \Phi = -\frac{\alpha}{2r}\left[1 + \frac{B^2}{r^2}\right].$$

In order to determine the equation of the orbit, Einstein could thus proceed exactly as in the Newtonian case. Einstein obtained from the above equation, using (8) and (8a) the relativistic law of conservation of energy for the orbit of the planet:[174]

$$\frac{dr^2 + r^2 d\phi^2}{ds^2} = 2A - 2\Phi = 2A + \frac{\alpha}{r} + \frac{\alpha B^2}{r^3}.$$

Einstein then eliminated s from this equation with the help of equation (10), and equated 1/r to x, and obtained the equation for the orbit of the planet,

$$(11)\ \left(\frac{dx}{d\phi}\right)^2 = \frac{2A}{B^2} + \frac{\alpha}{B^2}x - x^2 + \alpha x^3$$

Let us write this equation in a more familiar form:

$$(11a)\ \left[\frac{d}{d\phi}\left(\frac{1}{r}\right)\right]^2 = \frac{2E}{L^2} + \frac{2GM}{c^2L^2r} - \frac{1}{r^2} + \frac{\mathbf{2GM}}{\mathbf{c^2r^3}}.$$

Where again, x = 1/r, E is energy and L is the angular momentum (A = E and B = L).

This equation differs from the Newtonian equivalent equation in Newtonian theory only in the last term on the right-hand side. Einstein was now going to calculate that deviation.

Einstein integrated equation (11) and calculated the angle ϕ between the radius vector from the sun to the planet between the perihelion and the aphelion of the elliptical orbit of the planet, [175]

$$\phi = \int_{\alpha_1}^{\alpha_2} \frac{\mathrm{dx}}{\sqrt{\frac{2A}{B^2} + \frac{\alpha}{B^2}x - x^2 + \alpha x^3}},$$

(an elliptical integral) and $\alpha_1$ and $\alpha_2$ are the roots of the equation,

$$\frac{2A}{B^2} + \frac{\alpha}{B^2}x - x^2 + \alpha x^3 = 0.$$

Einstein wrote that $\alpha_1$ and $\alpha_2$ closely correspond to the roots of the Newtonian equation that arises from this equation by omission of $\alpha x^3$:

$$\frac{2A}{B^2} + \frac{\alpha}{B^2}x - x^2 = 0.$$

Thus Einstein considered now the integral without $\alpha x^3$, and he added to it the integral perturbing this Newtonian classical integral as a result of the presence of the term $\alpha x^3$. The result, using equation (8) and A = 0 (B = 1) in the above integral, was the following,

$$\phi = [1 + \alpha(\alpha_1 + \alpha_2)] \cdot \int_{\alpha_1}^{\alpha_2} \frac{\mathrm{dx}}{\sqrt{-(\mathrm{x} - \alpha_1)(\mathrm{x} - \alpha_2)(1 - \alpha x)}},$$

Finally the integration leads to:

$$\phi = \pi\left[1 + \frac{3}{4}\alpha(\alpha_1 + \alpha_2)\right].$$

The planet is orbiting the sun on an elliptical orbit. Einstein assumed that $\alpha_1$ and $\alpha_2$ signify the reciprocal values of the mean distance of the planet from the sun at the perihelion and the aphelion, respectively. Thus the angle ϕ between the radius vector from the sun to the planet between the perihelion and the aphelion is,

$$\phi = \pi\left(1 + \frac{3}{2}\alpha(\alpha_1 + \alpha_2)\right)$$

Or,

$$(12) \quad \phi = \pi\left(1 + \frac{3}{2}\frac{\alpha}{a(1-e^2)}\right).$$

Einstein thus arrived at the desired equation for the perihelion advance in the sense of motion after a complete orbit: [176]

$$(13) \quad \varepsilon = 3\pi\frac{\alpha}{a(1-e^2)}.$$

a is the semimajor axis, and e is the eccentricity.

Let us explain it in a simpler way. The Newtonian solution to the equation of the orbit is the following:

$$r = \frac{r_0}{1 + ecos\phi},$$

where, r is the distance from one of the foci of the ellipse and $r_0$ is a constant.

The relativistic solution of equation (11) is the following,

$$r = \frac{r_0}{1 + ecos(\phi - d\phi)}.$$

The difference between the two equations is the term –dϕ.

One can write the solution to this equation in the form of (13):

$$d\phi = 3\pi\frac{2GM}{c^2 a(1-e^2)}.$$

On introducing the orbital period T Einstein wrote equation (13) in the following form: [177]

$$(14)\ \varepsilon = 24\pi^2 \frac{a^2}{T^2 c^2 (1 - e^2)},$$

where c denotes the velocity of light in cm per sec.

Einstein concluded his scheme by saying, "The calculation yields, for the planet Mercury, an advance of the perihelion of 43" per century, while the astronomers indicated $45'' \pm 5''$ as the unexplained remainder between observations and the Newtonian theory. This means full compatibility".[178] A great triumph for Einstein's November theory.

On January 17, 1916, Einstein wrote Paul Ehrenfest.[179] Einstein told Ehrenfest, "Imagine my joy at the recognition of the feasibility of the general covariance and at the result that the equations yield the correct perihelion motion of Mercury. I was beside myself for a few days in joyous excitement".[180]

Abraham Pais wrote that later Einstein told Adrian Fokker that his discovery had given him palpitations of the heart. Pais said that what he told Wander Johannes de Hass was even more profoundly significant: "when he saw that his calculations agreed with the unexplained astronomical observations, he had the feeling that something actually snapped in him".[181]

Recall that Einstein calculated the deflection of a light ray near the sun from equations (4b) and (5). As a matter a fact, Einstein could calculate the deflection of a light ray by considering the orbit of the light ray. He could thus arrive at the solution by calculating the geodesic equation of a light ray passing near the sun, and thus he would have arrived at an equation very similar to equation (11),

$$(11b)\ \left(\frac{dx}{d\phi}\right)^2 = D^2 - x^2 + \boldsymbol{\alpha x^3}.$$

Again the additional term $\alpha x^3$ is the correction term upon the classical solution; and this clearly shows that the effect of the bending of a light ray is a relativistic effect, and not, for example, an effect caused by some phenomenon predicted by an emission theory. This additional term thus causes the light ray to deflect from a straight line in a strong gravitational field.

It is thus interesting that Einstein chose to solve the problem of the deflection of the light ray right at the very beginning of his paper, and did not come back to this problem once he obtained equation (11).

## 5 Schwarzschild responds to the November 18 Perihelion paper

### 5.1 Schwarzschild's Letter to Einstein

On December 22 Karl Schwarzschild, the director of the Astrophysical Observatory in Potsdam, wrote Einstein from the Russian front. Schwarzschild set out to rework Einstein's calculation in his paper of November 18 of the Mercury perihelion problem. There are two versions of the letter. In the draft version Schwarzschild wrote, "In order to familiarize myself with your gravitational theory, I set myself the task of solving completely, if possible, the problem you posed in the paper on Mercury's perihelion, and solved approximately. It is easy to specify the most general line element, which has the necessary information for symmetry properties. On this basis, I initially found following your way a certain first-order approximation,

$$g_{\rho\sigma} = -\alpha\frac{x_\rho x_\sigma}{r^3} - \beta\frac{x_\rho x_\sigma}{r^5} + \delta_{\rho\sigma}\left[\frac{\beta}{3r^3}\right], \quad g_{44} = 1$$

therefore, *two* arbitrary quantities α and β […]".[182]

[Recall Einstein's equation (4b) from the November 18 paper:[183]

$$(4b) \qquad g_{\rho\sigma} = -\delta_{\rho\sigma} - \alpha\frac{x_\rho x_\sigma}{r^3}, \quad g_{44} = 1 - \frac{\alpha}{r}.]$$

In the letter to Einstein, Schwarzschild omitted the first term (α = 0) in the above equation and wrote his approximate solution in a form similar to Einstein's solution,

$$g_{\rho\sigma} = -\frac{\beta x_\rho x_\sigma}{r^5} + \delta_{\rho\sigma}\left[\frac{\beta}{3r^3}\right], \quad g_{44} = 1.$$

Schwarzschild wanted to demonstrate that, the problem would be physically undetermined if there are a few approximate solutions.

Then Schwarzschild went over to the complete solution. He said he realized that there was only one line element, which satisfied Einstein's four conditions 1) to 4) imposed on the gravitational field of the sun [184] from his November 18 paper (static, spherical symmetric and tends to Minkowski's metric), as well as the field equations (1) from the November 18 paper; but Schwarzschild added that this line element is singular at the origin and only at the origin.

Schwarzschild considered a body, the origin of the coordinates is its geometric center. If one assumes isotropy of space and a static solution, which does not change with time, then there exists spherical symmetry around the center; and one can work with a system of spherical coordinates R, $\vartheta$, φ. The symmetry of the solution means that the variables are independent of the angular coordinates $\vartheta$, φ. Since the solution is static, there is no dependence on time, and thus only R is an independent variable, the distance from the center.

Schwarzschild originally wrote to Einstein the following,

"Let [the spherical coordinates],

$$x_1 = r\cos\varphi\cos\vartheta, \quad x_2 = r\sin\varphi\cos\vartheta, \quad x_3 = r\sin\vartheta,$$

[and a line element satisfying Einstein's field equations, the determinant condition and the 4 conditions on $g_{\mu\nu}$. Schwarzschild wrote this line element in Cartesian coordinates and with the above spherical coordinates wrote it in polar coordinates]. Consider,

$$R = (r^3 + \alpha^3)^{1/3} = r\left(1 + \frac{1}{3}\frac{\alpha^3}{r^3} + \cdots\right),$$

then the line element becomes [the Schwarzschild line-element]:

$$(4c)\ ds^2 = \left(1 - \frac{\gamma}{R}\right)dt^2 - \frac{dR^2}{1 - \frac{\gamma}{R}} - R^2(d\vartheta^2 + \sin^2\vartheta d\varphi^2)."$$

$$\sqrt{|-g|} = R^2\sin^2\vartheta\,.$$

Schwarzschild wrote that R, $\vartheta$, φ are not "allowed" coordinates, with which the field equations could be formed, because these spherical coordinates did not have determinant 1;[185] did not satisfy the coordinate condition (21b) from Einstein's November 11 paper, $\sqrt{-g} = 1$.[186]

Schwarzschild chose then the non-"allowed" coordinates, and in addition, a mathematical singularity was seen to occur in his solution when R = 0.

In the letter to Einstein Schwarzschild wrote that although the above coordinates did not have the determinant 1, the above line element expressed itself as the best in spherical coordinates.[187]

Schwarzschild added, "The equation of the orbit remains exactly as you obtained in the first approximation (11),[188] except that it must be understood that, for x not 1/r, but 1/R, which is a difference of the order of $10^{-12}$, thus practically absolutely irrelevant".[189]

Toward the end of his letter, Schwarzschild returned to his first approximation solution to Einstein's field equations. He wrote that the difficulty with the two arbitrary constants α and β, which the first approximation gave, is solved in that β must have a certain value of the order of $\alpha^3$, so that the way α is given, the solution be divergent by continuation of the approximation. [190]

Schwarzschild ended his letter to Einstein by saying that, "As you can see, the war is kindly with me, giving me fire, in spite of fierce gunfire, allowing in the very terrestrial distance, this stroll in your land of ideas".[191]

Earman and Janssen wrote Einstein's metric in equation (4b) in the form of a line element:

$$ds^2 = \left(1 - \frac{\alpha}{r}\right) \mathrm{dt}^2 - \sum_{\rho\sigma} \left(\delta_{\rho\sigma} + \alpha \frac{\mathrm{x}_\rho \mathrm{x}_\sigma}{\mathrm{r}^3}\right) d\mathrm{x}_\rho d\mathrm{x}_\sigma.$$

And this equation written in spherical coordinates becomes,

$$ds^2 = \left(1 - \frac{\alpha}{r}\right) dt^2 - \left(1 + \frac{\alpha}{r}\right) dr^2 - r^2(d\vartheta^2 + \sin^2\vartheta d\varphi^2),$$

which is the first order in α/r approximation to the exact Schwarzschild line element,

$$(4d)\ \ ds^2 = \left(1 - \frac{\alpha}{R}\right) dt^2 - \left(1 + \frac{\alpha}{R}\right)^{-1} dR^2 - R^2(d\vartheta^2 + \sin^2\vartheta d\varphi^2)$$

where, $\alpha = 2r = 2GM/c^2$ ($c^2 = 1$ here).

Einstein replied to Schwarzschild on December 29, 1915 and told him that his calculation proving uniqueness proof for the problem is very interesting. "I hope you publish the idem soon! I would not have thought that the strict treatment of the point-problem was so simple".[192]

### 5.2 Schwarzschild's paper

Subsequently Schwarzschild sent Einstein a manuscript, in which he derived his solution of Einstein's field equations for the field of a single mass.[193] Einstein received the manuscript by the beginning of January 1916, and he examined it "with great interest". He told Schwarzschild that he "did not expect that one could formulate so easily the rigorous solution to the problem".[194]

On the following Thursday, on January 13, Einstein delivered Schwarzschild's paper before the Prussian Academy with a few words of explanation. [195] Schwarzschild's paper, "Über das Gravitationsfeld eines Massenpunktes nach der Einsteinschen Theorie" (On the Gravitational Field of a Point-Mass according to Einstein's Theory) was published a month later in the *Sitzungsberichte der Königlich Preußischen Akademie der Wissenschaften.*

In the paper, Schwarzschild considered Einstein's field equations, the field equations (1) from the November 18 paper, and the "Determinantengleichung" (3). [196] Consider also the components of the gravitational field and Einstein's equations of the geodesic line (the equations of motion of a material point) – (15a) and (15b) of the November 4 paper.

Schwarzschild searched for a solution of the field equations satisfying Einstein's 4 conditions: the $g_{\mu\nu}$ are all independent of $x_4$, $g_{\rho 4} = g_{4\rho} = 0$, the solution is spherically symmetric, and it vanishes at infinity.

Then accordingly Schwarzschild wrote the line element satisfying Einstein's above four conditions in polar (spherical) coordinates (Polarkoordinaten):[197]

$$(6)\ ds^2 = Fdt^2 - (G + Hr^2)dr^2 - Gr^2(d\vartheta^2 + \sin^2\vartheta d\varphi^2),$$

Where F, G, h are functions of r.

Now the problem with Schwarzschild's spherical coordinates appeared:

$r^2 \sin^2\vartheta dr d\vartheta d\varphi$ is the volume element, and on performing transformations from the old coordinates to the new ones, the determinant $r^2 \sin^2\vartheta$ differs from 1.[198]

One could use these spherical coordinates but the transformations were complicated. However, Schwarzschild sitting in the Russian front, found a "simple trick" that allowed him to avoid this problem: "Ein einfacher Kunstgriff gestattet jedoch, diese Schwierigkeit zu umgehen".[199] And here exactly lies Schwarzschild's mathematical virtues and virtuoso, which Einstein lacked.

Schwarzschild assumed,

$$(7)\ x_1 = \frac{r^3}{3}, x_2 = cos\vartheta, x_3 = \varphi, x_4 = t\ .$$

Then,

$$r^2 dr sin\vartheta d\vartheta d\varphi = dx_1 dx_2 dx_3$$

holds in the whole volume element. "*The new variables are therefore spherical coordinates of the determinant 1*". In addition, Einstein's field equations (1) and the determinant equation (3) from his November 18 paper remain unchanged.[200]

Schwarzschild wrote (6) with the new spherical coordinates (7). From this he obtained:

$$(9)\ ds^2 = f_4 dx_4{}^2 - f_1 dx_1{}^2 - f_2 \frac{dx_2{}^2}{1 - x_2{}^2} - f_3 dx_3{}^2[1 - x_2{}^2].$$

Schwarzschild obtained three functions (in fact four, because two are equal according to Einstein's conditions): $f_1$, $f_2$, and $f_3$ are functions which satisfy Einstein's (3),

$f_1 f_2\ f_3\ f_4 = f_1 f_2^2\ f_4 = 1.$[201]

In addition, they satisfy:

For $x_1 = \infty,\ f_1 = \frac{1}{r^4} = (3x_1)^{-\frac{4}{3}},\ f_2 = f_3 = r^2 = (3x_1)^{\frac{2}{3}}, x_4 = 1.$

From (9) Schwarzschild obtained the components of the gravitational field and Einstein's field equations. Using Einstein's field equations, components of the gravitational field and the determinant condition (3), Schwarzschild obtained by integration, the following values for the functions $f_1$, $f_2$, and $f_4$:[202]

$$(10)\ f_2 = f_3 = (3x_1 + \rho)^{\frac{2}{3}},$$

$$(11)\ f_4 = 1 - \alpha(3x_1 + \rho)^{-\frac{1}{3}},$$

$$(12)\ f_1 = \frac{(3x_1 + \rho)^{-\frac{4}{3}}}{1 - \alpha(3x_1 + \rho)^{-\frac{1}{3}}}.$$

$\alpha$ and $\rho$ are two integration constants that remained arbitrary, and thus the problem was not fully defined physically.[203]

If $\alpha(3x_1 + \rho)^{-\frac{1}{3}} = 1$, and $3x_1 = \alpha^3 - \rho$, then $f_1$ is singular (a discontinuity of $f_1$ occurs). However, this is a mathematical singularity. If we move this singularity and locate it at $x_1 = 0$, then we get, $\rho = \alpha^3$. That is, we so-called solved the problem, and we find a relation between the two integration constants $\rho$ and $\alpha$.

Schwarzschild arrived at the following relation for a mass-point,

$$R = (3x_1 + \rho)^{\frac{1}{3}} = (r^3 + \alpha^3)^{\frac{1}{3}},$$

and in light of $\rho = \alpha^3$,

$R^3 = r^3 + \rho$.

$$f_1 = \frac{(r^3 - \rho)^{-\frac{4}{3}}}{1 - \alpha(r^3 + \rho)^{-\frac{1}{3}}}, \; f_4 = 1 - \alpha(r^3 - \rho)^{-\frac{1}{3}}, \; f_2 = f_3 = (r^3 - \rho)^{\frac{2}{3}}.$$

And up to second order Schwarzschild obtained,[204]

$$f_1 = \frac{1}{r^4}\left[1 + \frac{\alpha}{r} - 4/3\frac{\rho}{r^3}\right].$$

Inserting the value for R to the above functions according to the above definition, that is, $R = (r^3 + \alpha^3)^{1/3}$, Schwarzschild got,

$$f_1 = \frac{1}{R^4}\frac{1}{1 - \frac{\alpha}{R}}, f_2 = f_3 = R^2, f_4 = 1 - \frac{\alpha}{R},$$

then, substituting the above formulae for the functions in equation (9), and coming back to the "standard" (that is, non-"allowed") spherical coordinates, we arrive at the exact solution to Einstein's problem,

$$(14)\;\; ds^2 = \left(1 - \frac{\gamma}{R}\right)dt^2 - \frac{dR^2}{1 - \frac{\gamma}{R}} - R^2(d\vartheta^2 + \sin^2\vartheta d\varphi^2).$$ [205]

From now on the problem was simple. Consider a point moving along the geodesic line in the gravitational field, according to the line-element (14). Schwarzschild was looking for the equation of motion for that point. He assumed isotropy of space and that the motion of the planet was confined to the equatorial plane ($\vartheta = 90^0, d\vartheta = 0$), and he obtained three integrals,[206]

$$(15)\left(1-\frac{\alpha}{R}\right)\left(\frac{dt}{ds}\right)^2-\frac{1}{1-\alpha/R}\left(\frac{dR}{ds}\right)^2-R^2\left(\frac{d\phi}{ds}\right)^2=const.=h.$$

*h* is a constant of integration.

$$(16)\ R^2\frac{d\phi}{ds}=const.=c.$$

This equation is Einstein's equation (10).

This integral defines the time unit,

$$(17)\ (1-\alpha/R)\frac{dt}{ds}=const=1.$$

It then follows,

$$\left(\frac{dR}{d\phi}\right)^2+R^2(1-\alpha/R)=\frac{R^4}{c^2}[1-h(1-\alpha/R)].$$

The constant h = 1, and for x = 1/R one arrives at Einstein's equation (11):

$$\left(\frac{dx}{d\phi}\right)^2=\frac{1-h}{c^2}+\frac{h\alpha}{c^2}x-x^2-\alpha x^3,$$

where, $c^2/h = B$ (or c = B) and (1 – h)/h = 2A (A = 0), exactly as it is in equation (11) of Einstein's paper, and it gives the observed anomaly in the perihelion of Mercury.[207]

Later in 1916, David Hilbert approached Einstein's problem of solving the precession of the perihelion of Mercury and re-deriving Schwarzschild's line-element in his paper, "The Foundations of Physics (Second Communication)", presented to the Göttingen mathematical-physical class (on December 23, 1916).[208]

He followed Einstein's 3 conditions on the gravitational field of the sun, the solution is static, spherically symmetric, and $g_{r4} = g_{4r}$, but he did not follow Einstein's fourth condition, according to which $g_{\mu\nu}$ tends to Minkowskian values at infinity.[209]

Hilbert arrived at a line-element similar to that of Schwarzschild's one, which he designated as (45), and concluded that the singularity r = 0 disappears only if α = 0, "*the metric of the pseudo-Euclidean geometry is the only regular metric that corresponds to a world without electricity under the assumptions* 1., 2., 3".[210] Such an empty space was inacceptable by Einstein. If α ≠ 0, then r = 0; and for positive α also r = α. Thus another singularity exists, or as Hilbert puts it, these are places where the metric proves to be irregular. "The metric (45), [is] not regular at r = o and r = α is to be viewed as the expression for gravity of a centrally symmetric mass distribution in the neighborhood of the origin".[211] The singularity at r = α is not mathematical as the one at r = o.

**6. Hilbert and Einstein: Nostrification of my Theory**

Back to November 1915; after solving the problem of the Perihelion of Mercury Einstein could resolve the final difficulties in his November 11 theory. It probably took him an extra week to arrive at the November 25 Einstein tensor. On November 26, *a day after* Einstein presented his final version of the field equations, Einstein wrote his close friend Zangger,

"The general relativity problem is now finally dealt with. The perihelion motion of Mercury is explained wonderfully by the theory.

Astronomers have found from observations

$45'' \pm 5''$

I have found with the general theory of rel.

43".

Add to this the line shift for fixed stars which, as you know, has also been secured [by Freundlich in 1915], so this is already considerable confirmation of the theory. For the deflection of light by stars, this theory now provides an amount twice as large as before. I shall tell you verbally how this comes about.

The theory is beautiful beyond comparison. However, only *one* colleague has really understood it, and he is seeking to clearly "nostrify" it (Abraham's expression). In my personal experience I have hardly come to know the wretchedness of mankind sometimes better than this theory and everything connected to it. But it does not bother me".[212]

This colleague was David Hilbert. While Einstein was happy to have found in Hilbert one of the few colleagues, if not the only one, who appreciated and understood the nature of his work on gravitation, he also resented the way Hilbert took over some of his results without, as Einstein saw it, giving him due credit.[213]

In 1912, Max Abraham blamed Einstein's theory of relativity and indeed Einstein as well. Abraham thought that Einstein borrowed expressions from his new gravitation theory. However, it turned out that Abraham gradually converted to elements of Einstein's theory. Abraham needed Einstein's result of the mass of energy principle for his theory of gravitation. It was Abraham who corrected his theory according to Einstein's ideas, and not the other way round. However, after blaming Einstein, and because of his rejection towards relativity, Abraham could not accept this state of affairs. He thus found an original solution to the question, who actually arrived at the idea of the mass of energy?

Einstein explained all this to Ludwig Hopf on August 16, 1912, "Recently, Abraham – as you may have seen – slaughtered me along with the theory of relativity in two massive attacks, and wrote down (phys. Zeitschr.) the only correct theory of gravitation (under the 'nostrification' of my results) – a stately steed, that lacks three legs! He noted that the knowledge of the mass of energy comes from – Robert Mayer".[214]

Therefore, "nostrification" was Einstein's expression…

However, history repeats itself. First after the polemic with Abraham, Einstein encountered Gunnar Nördstrom. Like Abraham, Nördstrom had also gradually converted to Einstein's theory of gravitation. Nördstrom needed Einstein's corrections to his theory, his equivalence principle and also Einstein's help with conservation of energy (however, Nördstrom and Abraham also somewhat inspired Einstein).

In 1915 history repeated itself – once again. Hilbert was working on the same problem as Einstein. As seen up till now from Einstein's correspondence with Hilbert starting on November 7, it appears that Hilbert needed Einstein's November 4 and 1914 papers for his theory of gravitation. Hilbert could probably have not arrived at his theory without the help from Einstein. However, Hilbert also inspired Einstein in his November 11 and 18 Arbeits – as shown by Renn and Stachel.

Recall that on November 19 Hilbert sent Einstein a letter in which he congratulated him "on overcoming the perihelion motion". Hilbert ended his letter by asking Einstein to continue and keep him up to date on his latest advances. [215] Hilbert did not tell Einstein about the important talk he was giving the day afterwards. Hilbert presented on November 20 a paper to the Göttingen Academy of Sciences, "The Foundations of Physics", including the gravitational field equations of general relativity; the ones that Einstein presented to the Prussian Academy five days later on November 25. Could Einstein really take anything from this paper November 20 paper? In this section I am not going to examine Hilbert's work. For an extensive analysis of Hilbert's work consult Renn and Stachel.[216]

Since the Hilbert-Einstein priority dispute is very complicated, let us first start from Einstein's final presentation of November 25, and thereafter examine the "nostrification" issue from two points of view.

I will show here by a detailed examination of Einstein's November 25 paper that, it is reasonable to conjecture that Einstein could not have taken anything from Hilbert. *The main thesis that I present here is that Einstein's November 25 field equations sprang* *from his November 4 and his 1914 review article work. The entanglement between the two papers brought Einstein to his final solution of November 25. Einstein demonstrated this quite clearly in his review article of 1916.* It appears that after November 18 Einstein returned to his November 4 Weltanschauung.

Stachel says that after November 18 Einstein no more followed Hilbert's theory of matter.[217]

**6 Fourth Talk, November 25, 1915: Final Form of Field Equations**

On Thursday November 25 Einstein presented to the mathematical physical class of the Prussian Academy of sciences his short paper "Die Feldgleichungen der Gravitation" (The Field Equations of Gravitation").

**6.1 The History of the November Theory**

Einstein started his paper by briefly summarizing how in his previous two papers of November 4 and 11 he obtained field equations of gravitation that comply with the postulate of general relativity: i.e., field equations that in their general formulation are covariant under arbitrary substitutions of space-time variables.

Einstein explained that historically his field equations evolved in the following way.[218] First in the November 4 paper he found field equations

(16) $R_{\mu\nu} = -\chi T_{\mu\nu}$

which are covariant under arbitrary substitutions satisfying equation

$$(1)\quad \frac{\partial(x'_1 \ldots x'_4)}{\partial(x_1 \ldots x_4)} = 1.$$

Equations (16) contained the Newtonian theory:

$$(16b)\quad \frac{1}{2}\sum_{\alpha} \frac{\partial^2 g_{\mu\nu}}{\partial x_\alpha^2} = \chi T_{\mu\nu}$$

But (16) also led to:

$$(21a) \sum_{\alpha\beta} \frac{\partial}{\partial x_\alpha}\left(g^{\alpha\beta} \frac{\partial \lg\sqrt{-g}}{\partial x_\beta}\right) = -\chi \sum_{\sigma} T_\sigma^\sigma .$$

Einstein said that he found that (16) were equivalent to generally covariant field equations if the scalar of the energy tensor of "matter" vanishes. Then on November 11 he arrived at equations,

(16b)$_{\text{November 11}}$ $G_{\mu\nu} = -\chi T_{\mu\nu}$,

which could be specialized to (16) by introducing a new coordinate system in such a way that with respect to this system, $\sqrt{-g} = 1$ holds everywhere. This led to immense simplification of the equations.

In fact equations (16b)$_{\text{November 11}}$ for vacuum were at the basis of Einstein's third paper from November 18.

What happened between November 18 and 25 that brought Einstein back to the Prussian Academy with the final version of his field equations? Einstein wrote, "I now quite recently found that one can get along without this hypothesis about the energy tensor of matter, merely by inserting it into the field equations in a slightly different way. The field equations for vacuum, onto which I based the explanation of the Mercury perihelion, remain unaffected by this modification".[219]

Einstein summarized the important equations of his previous publications of November 4 and 11: [220]

$$(1)\ [(13)]\ G_{im} = R_{im} + S_{im}$$

$$(1b)\ [(13a)]\ R_{im} = -\sum_{l} \frac{\partial \begin{Bmatrix} im \\ l \end{Bmatrix}}{\partial x_l} + \sum_{l\rho} \begin{Bmatrix} il \\ \rho \end{Bmatrix} \begin{Bmatrix} m\rho \\ l \end{Bmatrix}$$

$$(1c)\ [(13b)]\ S_{im} = \sum_{l} \frac{\partial \begin{Bmatrix} il \\ l \end{Bmatrix}}{\partial x_m} - \sum_{l\rho} \begin{Bmatrix} im \\ \rho \end{Bmatrix} \begin{Bmatrix} \rho l \\ l \end{Bmatrix}.$$

The ten generally covariant equations of the gravitational field in space (16b)$_{\text{November 11}}$ of November 11 in the absence of matter are obtained by setting: [221]

(2) [(16b)$_{\text{November 11}}$] $G_{im} = 0$.

Einstein obtained,

$$(3)[(1)]\ R_{im} = \sum_{l} \frac{\partial \Gamma_{im}^{l}}{\partial x_l} + \sum_{\rho l} \Gamma_{i\rho}^{l} \Gamma_{ml}^{\rho} = 0$$

on choosing a reference system in which,

$$(3a)\ [(21b)]\ \sqrt{-g} = 1$$

$S_{im}$ vanishes, and equations (2) are simplified to the equations that Einstein used in his November 18 paper as the gravitational field of the sun.

Einstein has set as he had done in his November 4 paper,

$$(4)\ [(15)]\ \Gamma_{im}^{l} = -\begin{Bmatrix} im \\ l \end{Bmatrix}.$$

**6.2 The New Field Equation**

Subsequently Einstein wrote, [222]

"When 'matter' is present in the space under consideration, its energy tensor occurs on the right-hand sides of (2) and (3), respectively. We set:

$$(2a)\ \ G_{im} = -\chi\left(T_{im} - \frac{1}{2} g_{im} T\right),$$

where,

$$(5) \sum_{\rho\sigma} g^{\rho\sigma} T_{\rho\sigma} = \sum_{\sigma} T_{\sigma}^{\sigma} = T$$

is set. T is the scalar of the energy tensor of 'matter', the right-hand side of (2a) is a tensor".

Einstein again imposed the condition (3a) and obtained [see equation (3)] instead of (2a), [223]

$$(6)\ R_{im} = \sum_{l} \frac{\partial \Gamma_{im}^{l}}{\partial x_l} + \sum_{\rho l} \Gamma_{i\rho}^{l} \Gamma_{ml}^{\rho} = -\chi\left(T_{im} - \frac{1}{2} g_{im} T\right).$$

Einstein explained later to Besso on January 3, 1916, that his first paper (November 4) along with the addendum (November 11) still suffered from the lack of the second term on the right-hand side of this equation $\frac{1}{2} g_{im} T$ . Therefore the postulate T = 0 (from the addendum was necessary). But in his last paper no condition "results on the structure of matter" (and T = 0 is not necessary).[224] Einstein was going to demonstrate this in what followed.

Einstein wrote the linearized version of equation (6) already in 1912 in the Zurich Notebook on page 20L. In 1912 he extracted from the Ricci tensor linearized "gravitational equations":

$$\Sigma\ g_{\kappa\kappa} = \mathrm{U},$$

$$\Delta\left( g_{11} - \frac{1}{2} U\right) = \mathrm{T}_{11}, \quad \Delta\, g_{12} = \mathrm{T}_{12}, \ \ldots \quad \Delta\, g_{14} = \mathrm{T}_{14}.$$

The scholars did not find any indication in the notebook that Einstein tried to find the exact equations corresponding to these equations.[225] Did Einstein come back to page 20L before November 25, and could this give him the idea to extend his November 11 field equations to the November 25 field equations? There is no evidence supporting this scenario.

### 6.3 Conservation of Energy and Momentum

Einstein's next step in the November 25 paper was to prove conservation of energy-momentum. This also led him to explain the reasons for introducing the second term in (2a) and (6).

Einstein explained briefly to Sommerfeld on November 28:

"It is naturally easy to set these generally covariant equations down; however, it is difficult to recognize that they are generalizations of Poisson's equations, and not easy to recognize that they sufficiently satisfy the conservation laws".[226]

But Einstein realized that he could demonstrate that his field equation (2a) satisfied the conservation of momentum-energy, once he was "choosing the frame of reference so that $\sqrt{-g} = 1$. Then the equations take on the form,

$$[(6)] \quad -\sum_{l}\frac{\partial\begin{Bmatrix} im \\ l \end{Bmatrix}}{\partial x_l}+\sum_{\alpha\beta}\begin{Bmatrix} i\alpha \\ \beta \end{Bmatrix}\begin{Bmatrix} m\beta \\ \alpha \end{Bmatrix}=-\chi\left(T_{im}-\frac{1}{2}g_{im}T\right)$$

I had considered these equations with Grossman already 3 years ago, up to the second term on the right-hand side, but at that time had come to the conclusion that it did not lead to Newton's approximation, which was erroneous".[227]

Einstein specialized the choice of coordinated by (3a). Then $T_{im}$ satisfied the conditions, equation (14) times $g^{\mu\nu}$ of the November 4 paper, [228]

$$(7)\sum_{\lambda}\frac{\partial T_{\sigma}^{\lambda}}{\partial x_{\lambda}}=-\frac{1}{2}\sum_{\mu\nu}\frac{\partial g^{\mu\nu}}{\partial x_{\sigma}}T_{\mu\nu}$$

Einstein wrote this equation after equation (19a) in the November 4 paper [equation (19c)]. [229]

Einstein rewrote the first term on the right-hand side according to equation (15a) of his November 4 paper, exactly as he had done in this paper,

$$(7a)\sum_{\lambda}\frac{\partial T_{\sigma}^{\lambda}}{\partial x_{\lambda}}=-\sum_{\mu\nu}\Gamma_{\sigma\nu}^{\mu}T_{\mu}^{\nu}.$$

He then multiplied equations (6) by $\frac{\partial g^{im}}{\partial x_{\sigma}}$ and summed over i and m.

Therefore, because of (7) and (7a), [230]

$$\frac{1}{2}\sum_{im}g_{im}\frac{\partial g^{im}}{\partial x_{\sigma}}=-\frac{\partial lg\sqrt{-g}}{\partial x_{\sigma}}=0$$

[by (3a) $\sqrt{-g}=1$]

Einstein could prove conservation of energy-momentum for matter and gravitational field: [231]

$$(8)\ [(20)]\ \sum_{\lambda}\frac{\partial}{\partial x_{\lambda}}\left(T_{\sigma}^{\lambda}+\mathrm{t}_{\sigma}^{\lambda}\right)=0.$$

The stress-energy tensor of the gravitational field $\mathrm{t}_{\sigma}^{\lambda}$ was given by the corrected form of his $\mathfrak{L}$ [232] from November 4,[233]

$$(8a)\ \chi\mathrm{t}_{\sigma}^{\lambda}=\frac{1}{2}\delta_{\sigma}^{\lambda}\sum_{\mu\nu\alpha\beta}g^{\mu\nu}\Gamma_{\mu\beta}^{\alpha}\Gamma_{\nu\alpha}^{\beta}-\sum_{\mu\nu\alpha}g^{\mu\nu}\Gamma_{\mu\sigma}^{\alpha}\Gamma_{\nu\alpha}^{\lambda}.$$

Einstein said that $T_{im}$ of equation (6) should satisfy the condition (7) when restricted to the coordinate systems in which (3a) was valid. Equation (7) was equivalent to equation (14) from Einstein's November 4 paper: [234]

$$(14)\ \sum_{\nu}\frac{\partial T_{\sigma}^{\nu}}{\partial x_{\nu}}=\frac{1}{2}\sum_{\mu\tau\nu}g^{\tau\mu}\frac{\partial g_{\mu\nu}}{\partial x_{\sigma}}T_{\tau}^{\nu},$$

which came instead of equation (42a) from Einstein's 1914 paper.

It is possible that the term on the right-hand side of the above equation gave Einstein the idea for the second term (the scalar of the energy tensor) of equations (2a) and (6).

**6.4 No need for conditions on the Stress-Energy Tensor**

Einstein indeed explained the reasons for him adding the second term to equations (2a) and (6). Conservation of momentum-energy motivated him on the first place to add this term. This was in fact the exact motivation that led him to the "Entwurf" field equations in 1912 and to the rejection of the November tensor in that year.

Einstein contracted equation (6) – multiplied it by $g^{im}$ and summed over i and m. He obtained the following scalar equations,[235]

$$(9)\sum_{\alpha\beta}\frac{\partial^{2}g^{\alpha\beta}}{\partial x_{\alpha}\,\partial x_{\beta}}-\chi(T+t)=0.$$

When fully contracting (6) one obtains,

$$[(9a)]\sum_{\alpha\beta}\frac{\partial^2 g^{\alpha\beta}}{\partial x_\alpha \partial x_\beta} - \chi(T+t) + \sum_{\alpha\beta}\frac{\partial}{\partial x_\alpha}\left(g^{\alpha\beta}\frac{\partial \lg\sqrt{-g}}{\partial x_\beta}\right) = 0.$$

In correspondence with (5) Einstein used, [236]

$$(8b)\sum_{\rho\sigma} g^{\rho\sigma} t_{\rho\sigma} = \sum_{\sigma} t_\sigma^\sigma = t.$$

The additional term t in (9) represents the stress-energy tensor of the gravitational field, and it occurs in (9) on an equal footing with the stress-energy-tensor of matter. This, says Einstein, was not the case in equation (21) of his November 4 paper. Einstein derived another scalar equation: instead of (22) from his November 4 paper,[237]

$$(22)_{\mathrm{Nov4}}\ \frac{\partial}{\partial x_\mu}\left[\sum_{\alpha\beta}\frac{\partial^2 g^{\alpha\beta}}{\partial x_\alpha \partial x_\beta} - \chi t\right] = 0,$$

he wrote, [238]

$$(10)\frac{\partial}{\partial x_\mu}\left[\sum_{\alpha\beta}\frac{\partial^2 g^{\alpha\beta}}{\partial x_\alpha \partial x_\beta} - \chi(T+t)\right] = 0.$$

Hence,

$$\sum_{\alpha\beta}\frac{\partial^2 g^{\alpha\beta}}{\partial x_\alpha \partial x_\beta} - \chi(T+t) = 0.$$

Inserting (10) into (9a), gives instead of equation (21a) of November 4 the following equation,

$$[(11)]\sum_{\alpha\beta}\frac{\partial}{\partial x_\alpha}\left(g^{\alpha\beta}\frac{\partial \lg\sqrt{-g}}{\partial x_\beta}\right) = 0.$$

Einstein concluded that there was no need to impose other preconditions on the stress-energy tensor of matter, other than it complies with the theorem of energy-momentum[239] (that is, there was no need any more to assume the condition $\sum T_\mu^\mu = 0$).

Einstein could already possess equations (10) and (9) right after November 18, and at that time he very likely possessed the general idea of equation (6) that led to equations (9) and (10).

In a letter to Lorentz from January 19, 1916 Einstein explained that his field equations (2a) are the only equations that fulfill the following conditions:[240]

1) General covariance.
2) First order components $t_\mu^\nu$ of matter and derivations of the $g^{im}$'s higher than the second, do not appear in the $\frac{\partial^2 g^{im}}{\partial x_\alpha \partial x_\beta}$'s and the stress-energy tensor [equation (9)].
3) Compatibility with the "conservation law" of matter [(equation (8)] without any other restrictions for the $T_{mn}$'s.

Condition 3 was the conclusion that Einstein drew from equations (9) and (10) and finding that the second term was $\chi(T + t)$.

Einstein concluded his study by saying,[241]

"This then is finally the general theory of relativity completed as a logical structure. The postulate of relativity in its most general version, which makes space-time coordinates into physically meaningless parameters, leads with compelling necessity to a very specific theory of gravitation that explains the motion of the perihelion of Mercury".

On December 10, 1915 Einstein summarized the situation to his best friend Michele Besso,[242]

"Ich sandte Dir heute die Arbeiten . Die kühnsten Träume sind nun in Erfüllung gegangen. Allgemeine Kovarianz. Perihelbewegung des Merkur wunderbar genau. [...] Diesmal ist das Nächstliegende das Richtige gewesen; aber Grossmann u. ich glaubten, dass die Erhaltungssätze nicht erfüllt seien, und das Newton'sch Gesetz in erster Näherung nicht herauskomme. [...]

Herzliche Grüsse, [...] von Deinem zufriedenen aber ziemlich kaputen

Albert".

Einstein told Besso that his wildest dreams have now come true: general covariance and the perihelion of Mercury. This time the obvious was the right thing; but (it was not so obvious for Einstein before November 1915) Einstein and Grossmann believed that the conservation law was not satisfied by their theory; neither they believed they would obtain Newton's law to first approximation. Einstein wished Besso best regards (and also to his wife Anna Winteler and her father, with whom he stayed as a youth while studying in Aarau, and meanwhile became his relative), and signed "your satisfied kaput Albert".

**7 Einstein influences Hilbert**

Einstein's biographer Albrecht Fölsing wrote,

"Der 'nostrifizierende' Kollege war ausgerechnet David Hilbert, von dem er im Sommer noch 'ganz begeistert' war aber noch mehr dürfte ihn geärgert haben, daß Hilbert sogar schon einige Tage früher als er selbst die richtigen Feldgleichungen veröffentlicht hatte".[243]

The "nostrified" colleague (referred to in Einstein's letter to Zangger[244]) was Hilbert, and he was excited in the summer, a few days before Einstein had published the November 25 field equations. Einstein presented his field equations on November 25, 1915, but six days earlier, on November 20, Hilbert had derived the identical field equations for which Einstein had been searching such a long time. Fölsing asked about the possible draft that Hilbert sent Einstein before November 18, "Hat Einstein beim Uberfliegen dieser Arbeit denjenigen Term entdeckt, der in seinen Gleichungen noch fehlte, und dadurch etwa Hilbert 'nostrifiziert'?" Could Einstein, casting his eye over this paper, have discovered the term which was still lacking in his own equations, and thus 'nostrified' Hilbert?"[245]

Fölsing took Einstein's phraseology "nostrifiziert" and turned it against him. In fact, it is reasonable to assume that Einstein had discovered the second term in his equations (2a) by "casting his eye over" his own Arbeit of November 4, by reconsidering his equations of conservation of momentum-energy. This hypothesis is mostly strengthened by Einstein's own explanation from his November 25 paper, and the findings regarding Hilbert's November 20 paper seem to deny Einstein's nostrification.

According to these findings Einstein could have probably not have taken anything pertaining to the November 25 field equations from Hilbert's paper, or the summary sent to him. Of course on may ask whether there is a chance that this summary contained the correct explicit field equations in a part that is today missing? Stachel says there is always a chance, but a careful examination of a small missing part of the first proofs of Hilbert's November 20 paper shows that there is hardly 1/3 of a page room left for a few lines of text and equations; and all the evidence indicates that these lines did not include Hilbert's final form of the field equations.[246]

Indeed, the situation was quite the reverse: the question is whether Hilbert might have taken something from Einstein's 25 November paper, because it was quite possible for him to do so.[247]

The main point is the following. Hilbert submitted his paper to the Göttingen Academy of Sciences on November 20, 1915. Einstein's paper in which he gave the final form of his generally covariant field equations was submitted to the Prussian Academy of Sciences on November 25, 1915. Hilbert very likely sent to Einstein before November 18 a summary of his November 20 work.[248]

The November 20, 1915 proofs bear a printer's date stamp, "6 December 1915". The paper was not yet published.[249] According to Fölsing, the November 20, 1915 proofs of Hilbert paper are equivalent to Hilbert's printed paper and thus contain an equivalent version of equation (2a) or (6) of Einstein's November 25 paper. However, the November 20, 1915 paper did not contain a generally covariant theory. And thus represent Hilbert's states of work submitted on November 20. Hilbert started correcting his proofs only on December 6, 1915. Einstein's November 25, 1915 paper was published on December 2, 1915. Hilbert's paper was published only on March 31, 1916, and he had plenty of time to correct his November 20, 1915 paper according to Einstein's November Arbeits, including that of November 25, 1915. Hilbert indeed rewrote his proofs from November 20 sometime between December 1915 and March 1916.[250]

There are differences between the proofs and the printed version of the paper from March 1916. In the November 20 proofs Hilbert asserted that the theory he was developing could not be generally covariant. He based his assertion on a slightly more sophisticated version of Einstein's Hole Argument, and he cites Einstein's 1914 discussion of this argument. Hilbert was adopting Einstein's line of reasoning from 1913-mid 1915 just at the time that Einstein was abandoning it, in favor of a return to general covariance.[251]

In addition, in Hilbert's proofs of November 20 the gravitational field equations do not appear explicitly.[252] In the published version of March 1916 – after Einstein had published the final form of his field equations – the expression equivalent in form to Einstein's November 25 equations (2a) and (6) is written down explicitly. Corry, Renn and Stachel indeed claim that knowledge of Einstein's result [equations (2a) and (6)] may have been crucial to Hilbert's introduction of the second term in his equation, which was equivalent in form to these equations of Einstein.[253] And thus Einstein's papers helped Hilbert in putting his November 20 paper in a malleable form containing generally covariant field equations.

Finally, in the proofs of 6 December, Hilbert supplemented his reference to the gravitational potentials $g_{\mu\nu}$ in handwriting with the phrase "first introduced by Einstein".[254]

However, Renn and Stachel found lately that, in the December 6 proofs in handwriting Hilbert not only acknowledged Einstein's priority but attempted to placate him. In the introduction to his paper Hilbert wrote,[255]

"In the following – in the sense of the axiomatic method – I would like to develop, essentially from three axioms a ~~new~~ system of basic equations of physics, of ideal beauty, concerning, I believe, the solution of the problems presented".

Hilbert deleted the word "new", and this is a clear indication that he had read Einstein's November 25 paper and recognized that his own equations are formally equivalent (because of where the trace term occurs) to Einstein's.

On December 20, 1915 Einstein wrote Hilbert,[256]

"There has been certain resentment between us, the cause of which I do not want to analyze. I have fought against the associated feeling of bitterness, and this with complete success. I think of you again with unmixed kindness, and I ask you to try to do the same with me. It is objectively a shame when two real guys that have emerged from this shabby world do not give each other a little pleasure".

In the March 1916 printed version of his November 20 paper Hilbert added a reference to Einstein's November 25 paper and wrote, "the differential equations of gravitation that result are, as it seems to me, in agreement with the magnificent theory of general relativity established by Einstein in his last papers".[257]

On May, 27, 1916 Hilbert replied to Einstein's letter, and then only in response to a post card from Einstein with some questions about Hilbert's first paper on his new theory, which has just been published. Hilbert invited Einstein to visit Göttingen again and stay with him; but in spite of several invitations over the next few years, Einstein never came, perhaps because of Einstein's poor health in the last years of World War I. However, they continued to correspond over issues connected with Hilbert's paper.[258]

I am indebted to Prof John Stachel for his assistance and invaluable suggestions. It should be noted that the contents of this paper are the sole responsibility of the author.

## Endnotes

[1] Stachel, private conversation.

[2] For all the equations referring to Einstein, 1914b, see the first paper in the series of these papers, "From the Berlin 'Entwurf' Field Equation to the Einstein Tensor I: October 1914 until Beginning of November 1915.

[3] Einstein, 1914b, pp. 1041-1042.

[4] Einstein, 1914b, p. 1043.

[5] Einstein, 1914b, p. 1053.

[6] Einstein, 1915a, p. 781.

[7] Einstein, 1915a, p. 782; see the first paper, "From the Berlin 'Entwurf' Field Equation to the Einstein Tensor I: October 1914 until Beginning of November 1915.

[8] Einstein, 1915a, p. 783.

[9] Einstein, 1915a, p. 785.

[10] Renn and Stachel, 2007 in Renn (2007), p. 903.

[11] Einstein to Hilbert, November 7, 1915, *CPAE*, Vol. 8, Doc. 136.

[12] Stachel, 2005f (response to questions).

[13] Einstein, 1915b, p. 799; Renn and Stachel, 2007 in Renn (2007), p. 905.

[14] Einstein, 1915b, pp. 799-800.

[15] Einstein, 1915b, p. 800.

[16] Einstein to Hilbert, November 12, 1915, *CPAE*, Vol. 8, Doc. 139.

[17] Hilbert to Einstein, November 13, 1915, *CPAE*, Vol. 8, Doc. 140.

[18] Stachel, 2005f (response to questions).

[19] Einstein to Hilbert, November 15, 1915, *CPAE*, Vol. 8, Doc. 144.

[20] *CPAE*, Vol. 8, note 1, p. 202.

[21] Einstein to Hilbert, November 18, 1915, *CPAE*, Vol. 8, Doc. 148.

[22] Stachel, private converstaion.

[23] Hilbert to Einstein, November 19, 1915, *CPAE*, Vol. 8, Doc. 149.

[24] *CPAE*, Vol. 4, Doc. 14, pp. 1-14; *CPAE*, Vol. 4, "The Einstein-Besso Manuscript on the Motion of the Perihelion of Mercury", pp. 349-351.

[25] Earman and Janssen, 1993 in Earman, janssen and Norton (1993), pp. 142-143.

[26] *CPAE*, Vol. 4, "The Einstein-Besso Manuscript on the Motion of the Perihelion of Mercury", p. 346.

[27] *CPAE*, Vol. 4, Doc. 14, p. 1.

[28] *CPAE*, Vol. 4, Doc. 14, p. 6.

[29] *CPAE*, Vol. 4, "The Einstein-Besso Manuscript on the Motion of the Perihelion of Mercury", p. 349.

[30] *CPAE*, Vol. 4, "The Einstein-Besso Manuscript on the Motion of the Perihelion of Mercury", p. 350; *CPAE*, Vol. 4, Doc. 14, p. 8.

[31] *CPAE*, Vol. 4, Doc. 14, p. 8.

[32] *CPAE*, Vol. 4, Doc. 14, p. 9.

[33] *CPAE*, Vol. 4, Doc. 14, note 44, p. 377.

[34] *CPAE*, Vol. 4, Doc. 14, p. 14.

[35] *CPAE*, Vol. 4, Doc. 14, p. 14; *CPAE*, Vol. 4, "The Einstein-Besso Manuscript on the Motion of the Perihelion of Mercury", p. 351.

[36] Renn and Stachel, in Renn (2007), p. 907.

[37] Einstein, 1915c, p. 837.

[38] Einstein, 1915c, p. 839.

[39] Schwarzschild to Einstein, 22 December 1915, *CPAE*, Vol. 8, Doc. 169, note 5.

[40] Dokument Nr. 2 Protokoll der Sitzung der Phys.-math. Klasse der Akademie d. Wiss. vom 12 Juni 1913, Auszung, in Kirsten and Treder, 1979, p. 97.

[41] Einstein, 1915c, p. 833.

[42] Earman and Janssen, 1993 in Earman, Janssen and Norton (1993), pp. 144-145.

[43] Stachel, private conversation.

[44] Einstein, 1916, p. 820.

[45] Einstein to Zangger, November 26, 1915, *CPAE*, Vol. 8, Doc. 152.

[46] Stachel, private conversation.

[47] Einstein to Hans Albert Einstein, November 4, 1915, *CPAE*, Vol. 8, Doc. 134.

[48] Einstein, 1914b.

[49] Einstein to Hans Albert Einstein, November 4, 1915, *CPAE*, Vol. 8, Doc. 134.

[50] Einstein, 1914b, sections §12-§15, pages 1066-1077.

[51] Einstein, 1914b, pp. 1074-1076.

[52] Stachel, 1989 in Stachel (2002), p. 322.

[53] Stachel, 1989 in Stachel (2002), p. 322.

[54] Einstein to Lorentz, October 12, 1915, *CPAE*, Vol. 8, Doc. 129.

[55] Einstein, 1915a, p. 778.

[56] Einstein, 1915a, p. 778.

[57] Einstein, 1915a, p. 778.

[58] Einstein, 1915a, pp. 778-779.

[59] Stachel, private conversation.

[60] Einstein, 1915a, p. 779.

[61] Einstein, 1914b, pp. 1040-1041.

[62] Einstein, 1915a, p. 779.

[63] Einstein, 1915a, p. 779.

[64] Einstein, 1915a, p. 780.

[65] Einstein, 1915a, p. 780.

[66] Einstein, 1915a, p. 780.

[67] Einstein, 1915a, p. 780.

[68] Einstein, 1915a, p. 781.

[69] Einstein, 1915a, p. 781.

[70] Renn (2007), Vol. 2, pp. 451; 453.

[71] Einstein, 1915a, p. 782.

[72] Renn (2007), Vol. 2, pp. 451; 453.

[73] Einstein, 1915a, p. 782.

[74] Einstein, 1915a, p. 782.

[75] Einstein, 1915a, p. 782.

[76] Einstein, 1915a, p. 782.

[77] Einstein, 1915a, pp. 782-783.

[78] Stachel, 2007 in Renn (2007), p. 1058.

[79] Einstein to Sommerfeld, November 28, 1915, *CPAE*, Vol. 8, Doc. 153.

[80] Einstein to Lorentz, January, 1 1916, *CPAE*, Vol. 8, Doc. 177.

[81] Einstein, 1915a, p. 783.

[82] Einstein, 1915a, p. 783.

[83] Einstein, 1915a, p. 783.

[84] Einstein to Sommerfeld, November 28, 1915, *CPAE*, Vol. 8, Doc. 153.

[85] Einstein, 1915a, p. 784.

[86] Einstein, 1914b, p. 1073.

[87] Einstein, 1915a, p. 784.

[88] Einstein, 1915a, p. 784.

[89] Einstein, 1915a, p. 784.

[90] Einstein, 1914b, p. 1077.

[91] Einstein, 1914b, p. 1058.

[92] Einstein, 1914b, p. 1077.

[93] Einstein, 1915a, p. 785.

[94] Einstein, 1915a, p. 785.

[95] Einstein, 1915a, p. 785.

[96] Einstein, 1914b, p. 1070.

[97] Einstein, 1915a, p. 785.

[98] Einstein, 1915a, p. 785.

[99] Einstein, 1915a, p. 785.

[100] Einstein, 1915a, p. 786.

[101] Einstein, 1915a, p. 786.

[102] Janssen, Renn, Sauer, Norton and Stachel, 2007, in Renn (2007), p. 652; Renn (2007), Vol. 2, pp. 451; 453.

[103] Janssen, Renn, Sauer, Norton and Stachel, 2007, in Renn (2007), p. 554; Renn (2007), Vol. 2, pp. 454; 456.

[104] Einstein to Schwarzschild, February 19, 1916, *CPAE*, Vol. 8, Doc. 194, note 2.

[105] Einstein, 1915a, p. 786.

[106] Einstein to Hilbert, November 7, 1915, *CPAE*, Vol. 8, Doc. 136.

[107] Einstein to Hilbert, November 7, 1915, *CPAE*, Vol. 8, Doc. 136.

[108] *CPAE*, Vol. 8, note 4, p. 192.

[109] Einstein to Hilbert, March 30 1916, *CPAE*, Vol. 8, Doc. 207.

[110] Einstein, 1914b, p. 1073.

[111] Einstein, 1914b, p. 1073.

[112] Einstein to Hilbert, March 30 1916, *CPAE*, Vol. 8, Doc. 207.

[113] Einstein, 1915b, p. 799.

[114] Einstein, 1915b, pp. 799-800.

[115] Renn and Stachel, 2007, in Renn (ed), 2007, vol. 4, p. 905.

[116] Renn and Stachel, in Renn (ed), 2007, vol. 4, pp. 905-907.

[117] Einstein, 1915b, p. 800.

[118] Einstein, 1915b, p. 800.

[119] Einstein, 1915a, p. 780.

[120] Einstein, 1915b, p. 800.

[121] Einstein, 1915b, p. 801.

[122] Einstein, 1915b, p. 801.

[123] Einstein, 1915a, p. 785.

[124] Einstein to Hilbert, November 12, 1915, *CPAE*, Vol. 8, Doc. 139.

[125] Einstein to Hilbert, November 7, 1915, *CPAE*, Vol. 8, Doc. 136.

[126] Einstein to Hilbert, November 12, 1915, *CPAE*, Vol. 8, Doc. 139.

[127] Renn and Stachel, in Renn (2007), p. 904.

[128] Einstein, 1914b.

[129] Einstein to Hilbert, November 7, 1915, *CPAE*, Vol. 8, Doc. 136.

[130] Einstein to Hilbert, November 12, 1915, *CPAE*, Vol. 8, Doc. 139.

[131] Renn and Stachel, 2007, in Renn (2007), p. 869.

[132] Hilbert to Einstein, November 13, 1915, *CPAE*, Vol. 8, Doc. 140.

[133] Renn and Stachel, 2007, in Renn (207), p. 905.

[134] Hilbert to Einstein, November 13, 1915, *CPAE*, Vol. 8, Doc. 140.

[135] *CPAE*, Vol. 8, note 2, p. 196.

[136] Hilbert to Einstein, November 13, 1915, *CPAE*, Vol. 8, Doc. 140.

[137] Einstein to Hilbert, November 15, 1915, *CPAE*, Vol. 8, Doc. 144.

[138] *CPAE*, Vol. 8, note 1, p. 202.

[139] Einstein to Besso, 17 November, 1915, *CPAE*, vol 5, Doc 147; Renn and Stachel, 2007, in Renn (207), p. 906.

[140] Einstein to Hilbert, November 18, 1915, *CPAE*, Vol. 8, Doc. 148.

[141] Corry, Renn, Stachel, 2004, p. 6.

[142] Einstein, 1915a, p. 782.

[143] Hilbert to Einstein, November 19, 1915, *CPAE*, Vol. 8, Doc. 149.

[144] *CPAE*, Vol. 4, Doc. 14, pp. 1-14; *CPAE*, Vol. 4, "The Einstein-Besso Manuscript on the Motion of the Perihelion of Mercury", pp. 349-351.

[145] Renn and Stachel, in Renn (2007), p. 907.

[146] Einstein, 1915c, p. 831; Stachel, 1989 in Stachel (2002), p. 232.

[147] Einstein, 1915c, p. 831.

[148] Stachel, private converstaion.

[149] Einstein to Hilbert, November 18, 1915, *CPAE*, Vol. 8, Doc. 148.

[150] Einstein, 1915c, p. 832.

[151] Einstein, 1915a, p. 783.

[152] Einstein, 1915c, p. 832.

[153] Einstein, 1915c, p. 832.

[154] Earman and Janssen, 1993 in Earman, janssen and Norton (1993), p. 141.

[155] Einstein, 1915c, p. 832.

[156] Einstein, 1915c, p. 832.

[157] Earman and Janssen say that it is not clear whether Einstein meant here the metric field of the sun or the components of the gravitational field of the sun. Earman and Janssen, 1993 in Earman, janssen and Norton (1993), p. 142.

[158] Einstein, 1915c, p. 833.

[159] Einstein, 1915c, p. 833.

[160] Earman and Janssen, 1993 in Earman, janssen and Norton (1993), pp. 143-144.

[161] Einstein, 1915c, p. 833.

[162] Einstein, 1915c, p. 834.

[163] Einstein, 1915c, p. 834.

[164] Einstein, 1915c, p. 834.

[165] Einstein, 1915c, p. 835.

[166] Einstein, 1915c, p. 835.

[167] Einstein, 1915c, p. 835.

[168] Einstein, 1915c, p. 835.

[169] Einstein, 1915c, p. 836.

[170] Einstein, 1915c, p. 836.

[171] Einstein, 1915c, p. 837.

[172] Einstein, 1915c, p. 837.

[173] Einstein, 1915c, p. 837; in the original Einstein wrote this equation without a factor 1/r in the potential, see Earman and Janssen, 1993 in Earman, Janssen and Norton (1993), p. 152.

[174] Einstein, 1915c, p. 837.

[175] Einstein, 1915c, p. 838.

[176] Einstein, 1915c, p. 838.

[177] Einstein, 1915c, p. 839.

[178] Einstein, 1915c, p. 839.

[179] Einstein to Ehrenfest, January 17, 1916, *CPAE*, Vol. 8, Doc. 182; the exchange between Einstein and Ehrenfest will be examined later within the context of Einstein's review article of 1916.

[180] Einstein to Ehrenfest, January 17, 1916, *CPAE*, Vol. 8, Doc. 182.

[181] Pais, 1982, p. 253.

[182] Schwarzschild to Einstein, 22 December 1915, *CPAE*, Vol. 8, Doc. 169, note 5.

[183] Einstein, 1915c, p. 833.

[184] Einstein, 1915c, p. 833.

[185] Schwarzschild to Einstein, 22 December 1915, *CPAE*, Vol. 8, Doc. 169.

[186] Earman and Janssen, 1993 in Earman, Janssen and Norton (1993), pp. 144-145.

[187] Schwarzschild to Einstein, 22 December 1915, *CPAE*, Vol. 8, Doc. 169.

[188] Einstein, 1915c, p. 837.

[189] Schwarzschild to Einstein, 22 December 1915, *CPAE*, Vol. 8, Doc. 169.

[190] Schwarzschild to Einstein, 22 December 1915, *CPAE*, Vol. 8, Doc. 169, note 10.

[191] Schwarzschild to Einstein, 22 December 1915, *CPAE*, Vol. 8, Doc. 169.

[192] Einstein to Schwarzschild, 29 December 1915, *CPAE*, Vol. 8, Doc. 176.

[193] *CPAE*, Vol. 8, note 1, p. 242.

[194] Einstein to Schwarzschild, 9 January 1916, *CPAE*, Vol. 8, Doc. 181.

[195] Einstein to Schwarzschild, 9 January 1916, *CPAE*, Vol. 8, Doc. 181.

[196] Schwarzschild 1916a, p. 190.

[197] Schwarzschild 1916a, p. 191.

[198] Schwarzschild 1916a, p. 191.

[199] Schwarzschild 1916a, p. 191.

[200] Schwarzschild 1916a, p. 191.

[201] Schwarzschild 1916a, p. 191.

[202] Schwarzschild, 1916a, ppp. 193-194.

[203] Schwarzschild, 1916a, p. 195.

[204] Schwarzschild, 1916a, p. 194.

[205] Schwarzschild 1916a, p. 194.

[206] Schwarzschild 1916a, p. 195.

[207] Schwarzschild, 1916a, p. 195.

[208] Hilbert, 1917.

[209] Hilbert, 1917, p. 67; trans. Renn (ed), p. 1029.

[210] Hilbert, 1917, p. 70; trans. Renn (ed), p. 1033.

[211] Hilbert, 1917, p. 71; trans. Renn (ed), p. 1033.

[212] Einstein to Zangger, November 26, 1915, *CPAE*, Vol. 8, Doc. 152.

[213] Renn and Stachel, 2007 in Renn (2007), pp. 910-911.

[214] Einstein to Hopf, August 16, 1912, *CPAE*, Vol. 5, Doc 416.

[215] Hilbert to Einstein, November 19, 1915, *CPAE*, Vol. 8, Doc. 149.

[216] Renn and Stachel, 2007 in Renn (2007).

[217] Stachel, private converstaion.

[218] Einstein, 1915d, p. 844.

[219] Einstein, 1915d, p. 844.

[220] Einstein, 1915d, p. 844.

[221] Einstein, 1915d, p. 845.

[222] Einstein, 1915d, p. 845.

[223] Einstein, 1915d, p. 845.

[224] Einstein to Besso, January 3, 1916, *CPAE*, Vol. 8, Doc. 178.

[225] Janssen, Renn, Sauer, Norton and Stachel, 2007, in Renn (2007), p. 634.

[226] Einstein to Sommerfeld, November 28, 1915, *CPAE*, Vol. 8, Doc. 153.

[227] Einstein to Sommerfeld, November 28, 1915, *CPAE*, Vol. 8, Doc. 153.

[228] Einstein, 1915d, p. 845.

[229] Einstein, 1915a, p. 784.

[230] Einstein, 1915d, p. 846.

[231] Einstein, 1915a, p. 785.

[232] Einstein, 1915a, p. 784.

[233] Einstein, 1915d, p. 846.

[234] Einstein, 1915a, p. 782.

[235] Einstein, 1915d, p. 846.

[236] Einstein, 1915d, p. 846.

[237] Einstein, 1915a, p. 785.

[238] Einstein, 1915d, p. 846.

[239] Einstein, 1915d, pp. 846-847.

[240] Einstein to Lorentz, 19 January 1916, *CPAE*, Vol. 8, Doc. 184.

[241] Einstein, 1915d, p. 847.

[242] Einstein to Besso, Decmeber 10, 1915, *CPAE*, Vol. 8, Doc. 162.

[243] Fölsing, 1993, p. 420.

[244] Einstein to Zangger, November 26, 1915, *CPAE*, Vol. 8, Doc. 152.

[245] Fölsing, 1993, p. 421; Fölsing, 1999, pp. 375-376.

[246] Stachel, 2005f (response to questions).

[247] Stachel, 1999 in Stachel (2002), p. 357.

[248] Stachel, 1999 in Stachel (2002), p. 358.

[249] Corry, Renn and Stachel, 1997 in Stachel (2002), p. 340.

[250] Stachel, 1999 in Stachel (2002), p. 358.

[251] Stachel, 1999 in Stachel (2002), p. 359.

[252] Corry, Renn and Stachel, 1997 in Stachel (2002), p. 342; Stachel, 1999 in Stachel (2002), p. 359.

[253] Corry, Renn and Stachel, 1997 in Stachel (2002), p. 343.

[254] Corry, Renn and Stachel, 1997, in Stachel (2002), p. 344.

[255] Renn and Stachel, 2007 in Renn (2007), pp. 911-912.

[256] Einstein to Hilbert, December 20, 1915, *CPAE*, Vol. 8, Doc. 167.

[257] Corry, Renn and Stachel, 1997 in Stachel (2002), p. 344.

[258] Stachel, 2005f, p. 5.

-----------------------------------------------------------------------------------------------------

## References list

Corry, Leo, Renn, Jürgen and John Stachel, "Belated Decision in the Hilbert-Einstein Priority Dispute", 1997, in Stachel 2002, pp. 339-346.

Corry, Leo, Renn, Jürgen and John Stachel, "Response to F. Winterberg 'On Belated Decision in the Hilbert-Einstein Priority Dispute'", 2004, *Z. Naturforsen* 59a, pp. 715-719.

*The Collected Papers of Albert Einstein. Vol. 4: The Swiss Years: Writings, 1912–1914* (*CPAE* 4), Klein, Martin J., Kox, A.J., Renn, Jürgen, and Schulmann, Robert (eds.), Princeton: Princeton University Press, 1995.

*The Collected Papers of Albert Einstein. Vol. 5: The Swiss Years: Correspondence, 1902–1914* (*CPAE* 5), Klein, Martin J., Kox, A.J., and Schulmann, Robert (eds.), Princeton: Princeton University Press, 1993.

*The Collected Papers of Albert Einstein. Vol. 6: The Berlin Years: Writings, 1914–1917* (*CPAE* 6), Klein, Martin J., Kox, A.J., and Schulmann, Robert (eds.), Princeton: Princeton University Press, 1996.

*The Collected Papers of Albert Einstein. Vol. 7: The Berlin Years: Writings, 1918–1921* (*CPAE* 7), Janssen, Michel, Schulmann, Robert, Illy, Jószef, Lehner, Christoph, Buchwald, Diana Kormos (eds.), Princeton: Princeton University Press, 1998.

*The Collected Papers of Albert Einstein. Vol. 8: The Berlin Years: Correspondence, 1914–1918* (*CPAE* 8), Schulmann, Robert, Kox, A.J., Janssen, Michel, Illy, Jószef (eds.), Princeton: Princeton University Press, 2002.

*The Collected Papers of Albert Einstein. Vol. 9: The Berlin Years: Correspondence, January 1919–April 1920* (*CPAE* 9), Buchwald, Diana Kormos, Schulmann, Robert, Illy, Jószef, Kennefick, Daniel J., and Sauer, Tilman (eds.), Princeton: Princeton University Press, 2004.

Earman, John and Janssen, Michel, "Einstein's Explanation of the Motion of Mercury's Perihelion", in Earman, Janssen, and Norton, John (1993), pp. 129-172.

Ehrenfest, Paul, "Gleichformige Rotation starrer Korper und Relativitatstheorie", *Physikalische Zeitschrift* 10, 1909, pp. 918.

Einstein, Albert (1905a), "Zur Elektrodynamik bewegter Körper, *Annalen der Physik* 17, 1, 1905, pp. 891-921. (*CPAE* 2, Doc. 23). English translation in Einstein et al, (1952, pp. 35-65). (*CPAE* is Collected Papers of Albert Einstein)

Einstein, Albert, "Über das Relativitätsprinzip und die aus demselben gezogenen Folgerungen", *Jahrbuch der Radioaktivität* 4, pp. 411-462; 5, 1908, pp. 98-99 (Berichtigungen, errata). (*CPAE* 2, Doc. 47; 49).

Einstein, Albert (1911a), "Uber den Einfluβ der Schwerkraft auf die Ausbreitung des Lichtes", *Annalen der Physik* 35, 1911, pp. 898-908 (*CPAE* 3, Doc. 23). English translation in Einstein et al, (1952, pp. 99-108).

Einstein, Albert (1911b), "Zum Ehrenfestschen Paradoxon. Bemerkung zu V. Varičak's Aufsatz",. *Physikalische Zeitschrift* **12** (1911), pp. 509-510.

Einstein, Albert (1912a) Vol.1: *The Zurich Notebook and the Genesis of General Relativity* and Vol. 2:*Einstein's Zurich Notebook, Commentary and Essays* of J. Renn, ed., *The Genesis of General Relativity: Sources and Interpretation: Boston Studies in the Philosophy of Science* 250, 2007, Springer.

Einstein, Albert (1912b), "Lichtgeschwindigkeit und Statik des Gravitationsfeldes", *Annalen der Physik* 38, 1912, pp. 355-369 (*CPAE* 4, Doc. 3).

Einstein, Albert (1912c), "Zur Theorie des statischen Gravitationsfeldes", *Annalen der Physik* 38, 1912, pp. 443-458 (*CPAE* 4, Doc. 4).

Einstein, Albert (1912d), "Gibt es eine Gravitationswirkung, die der elektrodynamischen Induktionswirkung analog ist?", *Vierteljahrsschrift für gerichtliche Medizin und öffentliches Sanitätswesen* 44, pp. 37-40 (*CPAE* 4, Doc. 7).

Einstein, Albert (1912e), "Relativität und Gravitation. Erwiderung auf eine Bemerkung von M. Abraham", *Annalen der Physik* 38, 1912, pp. 1059-1064 (*CPAE* 4, Doc 8).

Einstein, Albert (1913a), "Zum gegenwärtigen Stande des Gravitationsproblems", *Physikalische Zeitschrift* 14, 1913, pp. 1249-1262 (*CPAE* 4, Doc. 17).

Einstein, Albert (1913b), "Physikalische Grundlagen einer Gravitationstheorie", *Vierteljahrsschrift der Naturforschenden Gesellschaft in Zürich*, 1914, pp. 284-290 (*CPAE* 4, Doc. 16).

Einstein, Albert, and Grossmann, Marcel, *Entwurf einer verallgemeinerten Relativitätstheorie und einer Theorie der Gravitation I. Physikalischer Teil von Albert Einstein. II. Mathematischer Teil von Marcel Grossman*, 1913, Leipzig and Berlin: B. G. Teubner. Reprinted with added "Bemerkungen", *Zeitschrift für Mathematik und Physik* 62, 1914, pp. 225-261. (*CPAE* 4, Doc. 13).

Einstein, Albert and Grossmann, Marcel, "Kovarianzeigenschaften der Feldgleichungen der auf die verallgemeinerte Relativitätstheorie gegründeten Gravitationstheorie", *Zeitschrift für Mathematik und Physik* 63, 1914, pp. 215-225.

Einstein, Albert (1914a), "Prinzipielles zur verallgemeinerten Relativitätstheorie und Gravitationstheorie", *Physikalische Zeitschrift* 15, 1914, pp. 176-180 (*CPAE* 4, Doc.25).

Einstein, Albert (1914b), "Die formale Grundlage der allgemeinen Relativitätstheorie", *Königlich Preußische Akademie der Wissenschaften* (Berlin). *Sitzungsberichte*, 1914, pp. 1030-1085 (*CPAE* 6, Doc. 9).

Einstein, Albert (1914c), "Zum Relativitäts-Problem", *Scientia (Bologna)* 15, 1914, pp. 337–348 (*CPAE* 4, Doc. 31).

Einstein, Albert (1914d), "Zur Theorie der Gravitation", *Naturforschende Gesellschaft, Zürich, Vierteljahrsschrift* 59, 1914, pp. 4–6 (*CPAE* 4, Doc. 27).

Einstein, Albert and Fokker, Adriann, D., "Die Nordströmsche Gravitationstheorie vom Standpunkt des absoluten Differentialkalküls", *Annelen der Physik* 44, 1914, pp. 321-328.

Einstein, Albert (1915a), "Zur allgemeinen Relativitätstheorie", *Königlich Preußische*, *Akademie der Wissenschaften* (Berlin). *Sitzungsberichte*, 1915, pp. 778-786 (*CPAE* 6, Doc. 21).

Einstein, Albert (1915b), "Zur allgemeinen Relativitätstheorie. (Nachtrag)", *Königlich Preußische Akademie der Wissenschaften* (Berlin). *Sitzungsberichte*, 1915, pp.799-801 (*CPAE* 6, Doc. 22).

Einstein, Albert (1915c), "Erklärung der Perihelbewegung des Merkur aus der allgemeinen Relativitätstheorie", *Königlich Preußische Akademie der Wissenschaften* (Berlin). *Sitzungsberichte*, 1915, pp. 831-839. (*CPAE* 6, Doc. 23).

Einstein, Albert (1915d), "Die Feldgleichungen der Gravitation", *Königlich Preußische Akademie der Wissenschaften* (Berlin). *Sitzungsberichte*, 1915, pp. 844-847 (*CPAE* 6, Doc. 25).

Einstein, Albert, "Die Grundlage der allgemeinen Relativitätstheorie", *Annalen der Physik* 49, 1916, pp. 769-822 (*CPAE* 6, Doc. 30). English translation in Einstein et al. (1952, pp. 111-164).

Einstein, Albert (1917a), *Uber die Spezielle und die Allgemeine Relativitätstheorie, Gemeinverständlich* ,1920, Braunschweig: Vieweg Shohn.

Einstein, Albert (1917b), "Kosmologische Betrachtungen zur allgemeinen", *Königlich Preußische Akademie der Wissenschaften* (Berlin). *Sitzungsberichte*, 1917, pp, 142–152. (*CPAE* 6, Doc. 43), English translation in Einstein et al. (1952, pp. 175-188).

Einstein, Albert (1918a), "über Gravitationwellen", *Sitzungsberichte der Königlish Preußischen Akademie der Wissenschaften* 1, 1918, pp. 154–167.

Einstein, Albert (1918b), "Prinzipielles zur allgemeinen Relativitätstheorie", *Annalen der Physik* 360, 1918, pp. 241-244.

Einstein, Albert (1918c), "Dialog über Einwände gegen die relativitästheorie", Die *Naturwissenschaften 6*, 1918, pp. 697-702.

Einstein, Albert (1920a), "Ather und Relativitätstheorie", 1920, Berlin: Springer. Reprinted in translation in Einstein (1983, pp. 1–24).

Einstein,Albert, (1920b) "Grundgedanken und Methoden der Relativitätstheorie in ihrer Entwicklung dargestellt", 1920, Unpublished draft of a paper for *Nature* magazine. (*CPAE* 7, Doc. 31).

Einstein, Albert, "Geometrie und Erfahrung, *Preußische Akademie der Wissenschaften* (Berlin). *Sitzungsberichte* I, 1921, pp.123-130. Reprinted in translation in Einstein (1983, pp. 27–56).

Einstein, Albert (1922a), Vier Vorlesungen Über Relativitätstheorie, 1922, Braunschweig: Vieweg, (*CPAE* 7, Doc. 71), Reprinted in translation as Einstein (1956).

Einstein, Albert (1922b), "How I Created the Theory of Relativity, translation to English by. Yoshimasha A. Ono, *Physics Today* 35, 1982, pp. 45-47.

Einstein, Albert, "Fundamental Ideas and problems of the Theory of Relativity", Lecture delivered on 11 July 1923 to the Nordic Assembly of Naturalists at Gothenburg, in acknowledgement of the Nobel Prize. *Reprinted in Nobel Lectures: Physics: 1901-1921*, 1967, New York: Elsevier, pp. 479-490.

Einstein, Albert, "Unpublished Opening Lecture for the Course on the Theory of Relativity in Argentina, 1925", translated by Alejandro Gangui and Eduardo L. Ortiz, *Science in Context* 21, 2008, pp. 451-459.

Einstein, Albert, *The Origins of the General Theory of Relativity*, 1933, Glasgow Jackson: Wylie & co (a booklet of 12 pages). Reprinted in slightly different version under the title "Notes on the Origin of the General Theory of Relativity" in Einstein 1954, pp. 285–290. This is a translation of the German version, "Einiges über die Entstehung der allgemeinen Relativitätstheorie", and is taken from Einstein 1934.

Einstein, Albert**,** *Mein Weltbild*, 1934, Amsterdam: Querido Verlag.

Einstein, Albert und Leopold, Infeld (1938a), *Physik als Abenteuer der Erkenntnis*, 1938, Leiden: A.W. Sijthoff. (Einstein's personal library, the Einstein Archives)

Einstein, Albert, and Infeld, Leopold (1938b), *The Evolution of Physics*, edited by Dr. C. P. Snow, 1938, The Cambridge library of Modern Science, Cambridge University Press, London. (Einstein's personal library, the Einstein Archives)

Einstein, Albert, "Consideration Concerning the Fundaments of Theoretical Physics", *Science* 91, May 24, 1940, pp. 487-492 (Address before the Eighth American Scientific Congress, Washington D.C., May 15, 1940). Reprinted as "The Fundaments of Theoretical Physics" in Einstein, 1954, pp. 323-335.

Einstein, Albert ,"Autobiographical notes" In Schilpp (1949, pp. 1–95).

Einstein, Albert und Infeld Leopold, *Die Evolution der Physik :von Newton bis zur Quantentheorie*, 1938/1956, Hamburg : Rowohlt. (Einstein's personal library, the Einstein Archives)

Einstein, Albert, *Zoals ik het zie: beschouwingen over maatschappelijke en wetenschappelijke onderwerpen, voor Nederland bewerkt en gecommentarieerd door,* 1951, H. Groot. Leiden: A. W. Sijthoff. (Einstein's personal library, the Einstein Archives) [in Dutch]

Einstein, Albert (1952a), *Aus Meinen Späten Jahren*, 1952, Zurich: Buchergilde Gutenberg.

Einstein, Albert (1952b), *Relativity, The Special and the General Theory*, Fifteenth Edition, trans. Robert W. Lawson, 1952, New-York: Crown Publishers, Inc.

Einstein, Albert, *Out of My Later Years*, 1956, New-York: Kensington Publishing Group.

Einstein, Albert**,** *Mein Weltbild*, edited by Carl Seelig, 1953, Zürich: Europa Verlag (with corrections by Einstein at the ETH Archives).

Einstein, Albert, *Ideas and Opinions*, 1954, New Jersey: Crown publishers (translated from Seelig's edition).

Einstein, Albert, "Erinnerungen-Souvenirs", *Schweizerische Hochschulzeitung* 28 (Sonderheft) (1955), pp. 145-148, pp. 151-153; Reprinted as, "Autobiographische Skizze" in Seelig, 1956, pp. 9-17.

Einstein, Albert, *Lettres à Maurice Solovine*, 1956, Paris: Gauthier Villars; *Letters to Solovine* (With an Introduction by Maurice Solovine, 1987), 1993, New York: Carol Publishing Group.

Einstein, Albert, *The Meaning of Relativity*, 5th ed, 1922/1956, Princeton: Princeton University Press. (Einstein's personal library, the Einstein Archives, 1922, 1945, 1950, 1953)

Einstein, Albert, *Relativity. The Special and the General Theory. A Clear Explanation that Anyone Can Understand*, 1959, New York: Crown Publishers.

Einstein, Albert and Besso, Michele, *Correspondence 1903-1955* translated by Pierre Speziali, 1971, Paris: Hermann.

Einstein, Albert, *Sidelights on Relativity*, 1983, New York: Dover.

Einstein, Albert, *Albert Einstein Max Born Brief Wechsel 1916-1955*, 1969/1991, Austria: Nymphenburger.

Einstein, Albert, *Letters to Solovine, with an Introduction by Maurice Solovine*, 1993, New-York: Carol Publishing Group.

Einstein, Albert, *Einstein's 1912 Manuscript on the Special Theory of Relativity*, 1996, Jerusalem: The Hebrew University of Jerusalem, Magnes.

Einstein, *Albert, Einstein's Masterpiece: The Foundation of General Relativity Die Grundlage der allgemeinen Relativitätstheorie The Foundations of the General Theory of Relativity Published May 11, 1916 in Annalen Der Physik*, 2010, Jerusalem: The Hebrew University of Jerusalem and the Academy of Sciences and Humanities of Israel.

Erikson, Erik, "Psychoanalytic Reflections of Einstein's Century", in Holton and Elkana, 1979/1997, pp. 151-171.

Frank, Philip, *Einstein: His Life and Times*, 1947, New York: Knopf, 2002, London: Jonathan, Cape.

Frank, Philip, *Albert Einstein sein Leben und seine Zeit*, 1949/1979, Braunschweig: F. Vieweg.

Frank, Philip, "Einstein's Philosophy of Science", *Reviews of modern Physics* 21, 1949, pp. 349-355.

Freundlich, Erwin, *Die Grundlagen der Einsteinschen Gravitationstheorie*, Mit einem Vorwort von Albert Einstein, 1916, Berlin: Verlag von Julius Springer.

Freundlich, Erwin Finlay, *The foundations of Einstein's theory of gravitation*, authorized English translation by Henry L. Brose; preface by Albert Einstein ; introduction by H.H. Turner, 1922, New York : G.E. Stechert. (Einstein's personal library, the Einstein Archives)

Grossman Marcel (1913) "Mathematische Begriffsbildungen zur Gravitationstheorie", *Vierteljahrsschrift der Naturforschenden Gesellschaft in Zürich*, 1914, pp. 291-297.

Glymour, Clark and John Earman "Relativity and Eclipses: The British Eclipse Expedition of 1919 and its Predecessors," *Historical Studies in the Physical Sciences* II, 1980, pp.49-85.

Goener, Hubert, Renn Jürgen, Ritter, Jim, Sauer, Tilman (ed), *The Expanding Worlds of General Relativity*, 1999, Boston: Brikhäser.

Hilbert, David, "Die Grundlagen der Physik. (Erste Mitteilung)", *Königliche Gesellschaft der Wissenschaften zu Göttingen. Mathematischphysikalische Klasse. Nachrichten*, 1915, pp. 395-407. Translation to English in Renn (2007) Vol. 4, pp. 1003-1015).

Hilbert, David, "Die Grundlagen der Physik (Zweite Mitteilung)", *Königliche Gesellschaft der Wissenschaften zu Göttingen mathematische physikalische Klasse, Nachrichten*, 1917, pp. 53-76. Translation to English in Renn (2007) Vol. 4, pp. 1017-1038).

Howard, Don, "Point Coincidences and Pointer Coincidences: Einstein on Invariant Structure in Spacetime Theories," in Goener et al (1999), pp. 463-500.

Howard, Don and Norton, John, "Out of the Labyrinth? Einstein, Hertz, and the Göttingen Answer to the Hole Argument", in Earman, Janssen, Norton (ed), 1993, pp. 30-61.

Howard, Don and Stachel, John (eds.), *Einstein and the History of General Relativity: Einstein Studies, Volume 1*, 1989, New York: Birkhauser.

Howard, Don and Stachel, John (eds.), *Einstein the Formative Years, 1879 – 1909: Einstein Studies, Volume 8*, 2000, New York: Birkhauser.

Janssen, Michel, "The Einstein-De Sitter Debate and its Aftermath", lecture, pp. 1-8, based on "The Einstein-De Sitter-Weyl-Klein Debate" in *CPAE*, Vol 8, 1998, pp. 351-357.

Janssen, Michel, "Rotation as the Nemesis of Einstein's *Entwurf* Theory", in Goener, Renn, Ritter and Sauer (ed), 1999, pp. 127-157.

Janssen, Michel, "The Einstein-Besso Manuscript: A Glimpse Behind the Certain of a Wizard", Freshman Colloquium: "Introduction to the Arts and Sciences", Fall 2002.

Janssen, Michel, "Of Pots and Holes: Einstein's Bumpy Road to General Relativity", in *Einstein's Annalen Papers. The Complete Collection 1901-1922*, editor, Jürgen Renn, 2005, Germany: Wiley-VCH, pp-58-85; reprinted as "Einstein's First Systematic Exposition of General Relativity", pp, 1-39.

Janssen Michel and Renn, Jürgen, "Untying the Knot: How Einstein Found His Way Back to Field Equations Discarded in the Zurich Notebook", in Renn (2007), Vol. 1, pp. 839-925.

Janssen, Michel, "What did Einstein know and When did he Know It?" in Renn (2007), Vol 2, pp. 786-837.

Janssen, Michel, "'No Success Like Failure': Einstein's Quest for General Relativity", *The Cambridge Companion to Einstein*, 2009 (not yet available in print).

Kirsten, Christa and Treder, Hans-Jürgen (ed), *Albert Einstein in Berlin 1913-1933, Teil I. Darstellung und Dokumente*, 1979, Berlin: Akademie-Verlag.

Klein, Martin, J., *Paul Ehrenfest, Volume 1 The making of a Theoretical Physicist*, 1970/1985, Amsterdam: North-Holland.

Kollros, Louis, "Erinnerungen eines Kommilitonen", 1955, in Seelig (1956), pp. 17-31.

Kramer, William, M., *A Lone Traveler, Einstein in California*, 2004, California: Skirball.

Kretschmann, Erich, "Über den physikalischen Sinn der Relativitätspostulate, A. Einstein neue und seine ursprügliche Relativitätstheorie", *Annalen der Physik* 53, 1917, pp. 575-614.

Krist, Jos. Dr., *Anfangsründe der Naturlehre für die Unterclassen der Realschulen*, 1891, Wien: Wilhelm Braumüller, K. U. K Hof. Und Universitäts-Buchhändler,

Laue, Max (1911a), *Die Relativitätstheorie*, *Erster band, das Relativitätsprinzip, der Lorentztransformation*, 1911/1921, Braunschweig: Vieweg & Sohn.

Laue Max (1911b) "Zur Diskussion über den starren Körporen in der Relativitätstheorie", *Physikalische Zeitschrift* 12, 1911, pp. 85-87.

Laue, von Max, "Zwei Einwände gegen die Relativitätstheorie und ihre Widerlegung", *Physikalische Zeitschrift* 13, 1912, pp. 118-120.

Laue, von Max, "Das Relativitätsprinzip", *Jahrbücher der Philosophie* 1, 1913, pp. 99-128.

Laue, von Max, *Die Relativitätstheorie, Zweiter Band, Die Allgemeine Relativitätstheorie und Einsteins Lehre von der Scwerkraft*, 1921, Germany: Braunschweig Druck und Verlag von Freidr. Vieweg & Sohn.

Laue von Max, *Gesammelte Schriften und Vorträge*, 1961, Berlin: Friedr. Vieweg & Sohn.

Levi-Civita, T., and Ricci-Curbastro, G., "Méthodes de calcul différential absolu et leurs applications", *Mathematische Annalen* 54 (1900), pp. 125–201

Levi-Civita, Tullio, *Der absolute Differentialkalkül und seine Anwendungen in Geometrie und Physik*, 1928, Berlin: Springer-Verlag. (Einstein's personal library, the Einstein Archives)

Levi-Civita, Tullio and Enrico Persico, *Lezioni di calcolo differenziale assoluto:*

*raccolte e compilate*, 1925, Roma: Alberto Stock. (Einstein's personal library, the Einstein Archives)

Levinson, Horace C., Zeisler Ernest Bloomfield, *The law of gravitation in relativity*, 1931, Chicago: The University of Chicago Press. (Einstein's personal library, the Einstein Archives)

Levinson, Thomas, *Einstein in Berlin*, 2003, New York: Bantam Books.

Mach, Ernest, *Mechanics: A Critical and Historical Account of Its Development*, translated by T. J. McCormack, 1893/1960, La Salle: Open court.

Mach, Ernest, *Popular Scientific Lectures*, 1894/1943, Ill: La Salle.

Mie, Gustav, "Grundlagen einer Theorie der Materie", *Annalen der Physik* 37, 1912, pp. 511–534; English translation in Renn (2007), pp. 633-697.

Mie, Gustav, "Bemerkungen zu der Einsteinschen Gravitationstheorie. I und II." *Physikalische Zeitschrift* 14, 1914, pp. 115–122, 169–176; English translation in Renn (2007), pp. 698-728.

Minkowski, Hermann (1907/1915), "Das Reltivitätsprinzip (Presented in Göttingen on 5.11.1907, Published Post-mortem by Arnold Sommerfeld), *Annalen der Physik* 47, 1915, pp. 927-938.

Minkowski, Hermann (1908a), "Die Grundgleichungen für die elektromagnetischen Vorgänge in bewegten Körpern, *Nachrichten von der Königlichen Gesellschaft der Wissenschaften zu Göttingen*, 1908, pp. 53-111.

Minkowski, Hermann (1908b), "Raum und Zeit" (lecture delivered on the 80$^{th}$ assembly of German Natural Scientists and Physicians, at Cologne on 21 September 1908), *Physikalische Zeitschrift* 20, 1909, pp.104-111.

Møller, Christian, *The Theory of Relativity* (Based on lectures delivered at the University of Copenhagen), 1952, Oxford: Clarendon Press. (Einstein's personal library, the Einstein Archives)

Moszkowski, Alexander (1921a), *Einstein, Einblicke in seine Gedankenwelt. Gemeinverständliche Betrachtungen über die Relativitätstheorie und ein neues Weltsystem. Entwickelt aus Gesprächen mit Einstein*, 1921, Hamburg: Hoffmann und Campe/ Berlin: F. Fontane & Co.

Moszkowski, Alexander (1921b), *Einstein the Searcher His Works Explained from Dialogues with Einstein*, 1921, translated by Henry L. Brose, London: Methuen & Go. LTD; appeared in 1970 as: *Conversations with Einstein*, London: Sidgwick & Jackson, 1970.

Newton, Isaac, *The Principia. Mathematical Principles of Natural Philosophy*, translated by Andrew Motte, 1726/1995, New York: Prometheus Books.

Nordmann, Charles, "Einstein Expose et Discute sa Théorie", *Revue des Deux*

*Mondes*, 1922, Tome Neuvième, Paris, pp. 130-166.

Nordström, Gunnar, "Relativitätsprinzip und Garavitation", *Physikalische Zeitschrift* 13, 1912, pp.1126-1129; English translation in Renn (ed), 2007, pp. 489-497.

Nordström, Gunnar (1913a), "Träge und schwere Mass in der Relativitätsmechanik", *Annalen der Physik* 40, 1913, pp. 856-878; English translation in Renn (ed), 2007, pp. 499-521.

Nordström, Gunnar (1913b), "Zur Theorie der Gravitation vom Standpunkt des Relativitätsprinzips", *Annalen der Physik* 42, 1913, pp. 533-554; English translation in Renn (ed), 2007, pp. 523-542.

Norton, John, "How Einstein Found His Field Equations: 1912-1915," *Historical Studies in the Physical Sciences* 14, 1984, pp. 253-315. Reprinted in D. Howard and J. Stachel (eds.), *Einstein and the History of General Relativity: Einstein Studies* Vol. I, Boston: Birkhauser, pp 101-159.

Norton, John, "General Covariance and the Foundations of General Relativity: Eight Decades of Dispute," *Reports on Progress in Physics* 56, 1993, pp.791-858.

Norton, John, "Einstein and Nordström: Some Lesser-Known Thought Experiments in Gravitation", *Archive for History of Exact Sciences* 45, 1993, pp.17-94.

Norton, John, "Nature in the Realization of the Simplest Conceivable Mathematical Ideas: Einstein and the Canon of Mathematical Simplicity", *Studies in the History and Philosophy of Modern Physics* 31, 2000, pp.135-170.

Norton, John, "Einstein, Nordström and the early Demise of Lorentz-covariant, Scalar Theories of Gravitation," Reprinted in revised form in Renn (ed), 2007, Vol. 3, pp. 413-487.

Pais, Abraham, *Subtle is the Lord. The Science and Life of Albert Einstein*, 1982, Oxford: Oxford University Press.

Pauli, Wolfgang (1921/1958), "Relativitätstheorie" in *Encyklopädie der Mathematischen wissenschaften*,Vol. 5, Part 2, 1921, Leipzig: Teubner, p. 539; English translation, *Theory of Relativity*, 1958, Oxford and New York: Pergamon.

Plesch János, *Die Herzklappenfehler einschließlich der allgemeinen Diagnostik, Symptomatologie und Therapie der Herzkrankheiten*, in Spezielle Pathologie und Therapie innerer Krankheiten, hrsg. von F. Kraus u.a., 1919-1927, Berlin etc: Urban und Schwarzenberg. (Einstein's personal library, the Einstein Archives)

Plesch, Johann, *Physiology and pathology of the heart and blood-vessels*,1937, London, Humphrey Milford, Oxford university press. (Einstein's personal library, the Einstein Archives)

Plesch, János, *János. Ein Arzt erzählt sein Leben*, 1949, Paul List Verlag, München / Leipzig.

Plesch, John, *János, The Story of a Doctor*, translated by Edward Fitzgerald, 1949,

New York, A.A. WYN, INC (first published 1947).

Reid, Constance, *Hilbert*, 1970/1996, New-York: Springer-Verlag.

Reiser, Anton [Rudolf Kayser] *Albert Einstein: A Biographical Portrait*, 1930/1952, New York : Dover. (Einstein's personal library, the Einstein Archives)

Reiser, Anton [Rudolf Kayser], *Albert Einstein, Ein Biographisches Porträt*, 1930/1997, New-York: Albert & Carles Boni.

Renn, Jürgen and Tilman Sauer, "Heuristics and mathematical Representation in Einstein Search for a Gravitational Field Equation", Preprint 62, *Max Planck Institute for the History of Science*, 1997.

Renn, Jürgen, (ed.) (2005a) *Einstein's Annalen Papers. The Complete Collection 1901-1922*, 2005, Germany: Wiley-VCH Verlag GmbH& Co.

Renn, Jürgen (2005b), *Albert Einstein, Chief Engineer of the Universe*, 2005, Berlin: Wiley-VCH.

Renn, Jürgen, "The Summit Almost Scaled: Max Abraham as a Pioneer of a Relativistic Theory of Gravitation", in Renn (ed), 2007, pp. 305-330.

Renn, Jürgen, Norton, John, Janssen, Michel and Stachel John, ed., *The Genesis of General Relativity*. 4 Vols., 2007, New York, Berlin: Springer.

Renn, Jürgen and Stachel, John , "Hilbert's Foundation of Physics: From a Theory of Everything to a Constituent of General Relativity", in Renn (ed), Vol. 4, pp. 857-974.

Renn, Jürgen, ed., *The Genesis of General Relativity*, Vol. 3 *Gravitation in the Twighlight of Classical Physics: Between Mechanics, Field Theory, and Astronomy*, 2007, New York, Berlin: Springer.

Robb, Alfred, *A Theory of Time and Space*, 1914, Cambridge: Cambridge University Press. (Einstein's personal library, the Einstein Archives)

Robinson, Andrew (ed), *Einstein A Hundred Years of Relativity*, 2005, New-York: Harry N. Abrams, Inc.

Russell, Bertrand, *Our knowledge of the external world*, 1926, London: Allen & Unwin. (Einstein's personal library, the Einstein Archives)

Samuel, Herbert Louis Samuel, Viscount, *Essay in physics*, with a letter from Albert Einstein, 1951, Oxford: B. Blackwell. (Einstein's personal library, the Einstein Archives)

Schilpp, Paul Arthur (ed.), *Albert Einstein: Philosopher-Scientist*, 1949 Library of Living Philosophers, Northwestern University, Evanston, ILL/ La Salle, IL: Open Court. (Einstein's personal library, the Einstein Archives)

Schwarzschild, Karl (1916a), "Über das Gravitationsfeld eines Massenpunktes nach der Einsteinschen Theorie", *Königlich Preußische Akademie der Wissenschaften (Berlin). Sitzungsberichte*,1916, pp.189-196.

Schwarzschild, Karl (1916b), "Über das Gravitationsfeld einer Kugel aus inkompressibler Flüssigkeit nach der Einsteinschen Theorie", *Königlich Preußische Akademie der Wissenschaften (Berlin). Sitzungsberichte*,1916, pp.424-434, wiederabgedruckt [reprinted], pp. 457-467.

Seelig, Carl (1952a), *Albert Einstein Und Die Scweiz*, 1952, Zürich: Europa Verlag.

Seelig Carl (1952b), *Albert Einstein; eine dokumentarische Biographie*, 1952, Zürich: Europa Verlag.

Seelig Carl, *Albert Einstein; eine dokumentarische Biographie*, 1954, Zürich: Europa Verlag.

Seelig Carl, *Albert Einstein: A documentary biography*, Translated to English by Mervyn Savill 1956, London: Staples Press.

Seelig Carl, *Helle Zeit – Dunkle Zeit. In memoriam Albert Einstein*, 1956, Zürich: Branschweig: Friedr. Vieweg Sohn/Europa.

Smeenk, Christopher and Martin Christopher, "Mie's Theories of Matter and Gravitation" in Renn (2007), pp. 623-632.

Stachel, John (1979a) "Einstein's odyssey: his Journey from Special to General Relativity", *The Sciences*, March 1979; reprinted in Stachel (2002), pp. 225-232.

Stachel, John (1979b), "The Genesis of General Relativity", *Physics* 100, 1979, pp. 428-442; reprinted in Stachel (2002), pp. 233-244.

Stachel, John, "The Rigidity Rotating Disk as the 'Missing Link' in the History of General Relativity", *General Relativity and Gravitation one Hundred Years After the Birth of Albert Einstein*, Vol 1, 1980, pp. 1-15; reprinted in Stachel (2002), pp. 245-260.

Stachel, John (1982a), "'Subtle is the Lord'"… The Science and Life of Albert Einstein" by Abraham Pais", *Science* 218, 1982, pp. 989-990; reprinted in Stachel (2002), pp. 551-554.

Stachel, John (1982b), "Albert Einstein: The Man Beyond the Myth", *Bostonia Magazine* 56, 1982, pp. 8–17; reprinted in Stachel (2002), pp. 3-12.

Stachel, John, "Einstein and the 'Research Passion'", Talk given at the Annual Meeting of the American Associates for the Advancement of Science in Detroit, May 1983; reprinted in Stachel (2002), pp. 87-94.

Stachel, John, "The Generally Covariant Form of Maxwell's Equations", in Earman, John, Janis, Allen, Massey, Gerald. I., and Rescher, Nicholas, *Philosophical Problems of the Inertial and External Worlds*, 1984, University of Pittbsbutgh/ Universitätswerlag Konstanz.

Stachel, John, "What a Physicist Can Learn From the Discovery of General Relativity", *Proceedings of the Fourth Marcel Grossmann Meeting on General relativity*, ed. R. Ruffini, Elsevier, 1986, pp.1857-1862.

Stachel, John (1987), "How Einstein Discovered General Relativity: A Historical Tale With Some Contemporary Morals", in MacCallum, M.A.H., *General Relativity and Gravitation Proceedings of the 11th International Conference on General Relativity and Gravitation*, 1987, pp. 200-209, reprinted in Satchel (2002), pp. 293-300.

Stachel, John, "Einstein's Search for General Covariance 1912-1915", in Howard, Don and Stachel John (eds), *Einstein and the History of General Relativity*, Einstein Studies, Vol 1, 1989, Birkhäuser, pp. 63-100; reprinted in Stachel (2002), pp. 301-338.

Stachel, John (1993a), "the Meaning of General Covariance. The Hole Story", in *philosophical problems of the Internal and external Worlds, Essays of the philosophy of Adolf Grünbaum*, Earman, John, Janis, Allen, I., Massey, Gerald, J., and Rescher, Nicholas (eds), 1993, Pittsburgh: University of Pittsburgh/Universitätsverlag Konstanz.

Stachel, John (1993b), "The Other Einstein: Einstein Contra Field Theory", *Science in Context* 6, 1993, pp. 275-290; reprinted in Stachel (2002), pp. 141-154.

Stachel, John (1994a), "Scientific Discoveries as Historical Artifacts" in Gavroglu, Kostas, Christianidis, Jean, and Nicolaidis, Efthymios (eds), *Trends in the Historiography of Science*, 1994, Boston/London/Dordrecht: Kluwer Academic.

Stachel, John (1994b), "Changes in the Concept of Space and Time Brought About by Relativity", *Artifacts, Representations and Social Practice: Essays for Marx Wartofsky*, eds Carol Gould and Robert S. Cohen., Dordrecht/Boston/London: Kluwer Academic, pp. 141-162.

Stachel, John, "History of relativity," in Brown, Laurie M., Pais, Abraham, and Sir Pippard, Brian (eds.), *Twentieth century physics*, 1995, New York: American Institute of Physics Press, pp. 249-356.

Stachel, John, "Albert Einstein: A Biography by Albert Fölsing", Review of Albrecht Fölsing, *Albert Einstein: A Biography*, in *Physics Today*, January, 1998; reprinted in Stachel (2002), pp. 555-556.

Stachel, John, "New Light on the Einstein-Hilbert Priority Question", *Journal of Astrophysics* 20, 1999, pp. 90-91; reprinted in Stachel (2002), pp. 353-364.

Stachel, John, "The First-two Acts", 2002/2007, in Renn, Jürgen et all (ed) (2007); reprinted first in Stachel (2002), pp. 261-292.

Stachel, John, *Einstein from 'B' to 'Z'*, 2002, Washington D.C.: Birkhauser.

Stachel, John(2005a), "Albert Einstein: A Man for the Millennium," reprinted in *Albert Century International Conference Paris, France 18-22 July 2005*, AIP Conference Proceedings 861, 2006, New York: American Institute of Physics.

Stachel, John (2005b), "Where is Creativity? The example of Albert Einstein", Invited Lecture at the Congresso International de Filosophia, Pessoa & Sociadade (Person and Society), Braga, 17-19 November 2005, unpublished.

Stachel, John (2005c), "Einstein and Hilbert", invited lecture in March 21, 2005 Session: Einstein and Friends, American Physical Society, Los Angeles; and response to Questions from FAZ on Hilbert and Einstein". Unpublished.

Stachel, John (2005d), "January 1905", Nature Supplement, *Nature*, vol 433, 2005, p. 258.

Stachel, John (2005e) "Einstein's Clocks, Poincare's Maps/Empires of Time", *Studies in History and Philosophy of Science* B 36 (1), pp. 202-210; pp. 1-15.

Stachel, John (2005f), "Albert Einstein: A Man for the Millennium," reprinted in *Albert Century International Conference Paris, France 18-22 July 2005*, AIP Conference Proceedings 861, 2006, New York: American Institute of Physics.

Stachel, John (2005g), "Albert Einstein", The 2005 Mastermind Lecture, *Proceedings of the British Academy* 151, 2007, pp.423-458.

Stachel, John (2005h), "Where is Creativity? The example of Albert Einstein", Invited Lecture at the Congresso International de Filosophia, Pessoa & Sociadade (Person and Society), Braga, 17-19 November 2005, unpublished.

Stachel, John (2005i), "Einstein and Hilbert", invited lecture in March 21, 2005 Session: Einstein and Friends, American Physical Society, Los Angeles; and response to Questions from FAZ on Hilbert and Einstein". Unpublished.

Stachel, John, "Einstein's Intuition and the Post-Newtonian Approximation", *World Scientific*, 2006, pp. 1-15.

Stachel, John (2007a), "The Story of Newstein: Or is Gravity Just Another Pretty Force?", in Renn, Jügen and Schimmel, Matthias (eds.), *The Genesis of General Relativity* vol. 4, *Gravitation in the Twilight of Classical Physics: The Promise of Mathematics* , 2007, Berlin, Springer, pp. 1041-1078.

Stachel, John (2007b), "Einstein", Master-Mind Lecture, *Proceedings of the British Academy* 151, 2007, pp. 423-458.

Stachel, John (2007c), "A world Without Time: the Forgotten Legacy of Gödel and Einstein", *Notices of the American Mathematical Society* 54, 2007, pp. 861-868 (1-8).

Stachel, John, "Albert Einstein", *The New Dictionary of Scientific Biography*, Vol 2, Gale 2008, pp. 363-373.

Stachel, John, "The Hole Argument", *Living Reviews in Relativity*, June 25, 2010.

Stachel, John, "The Rise and Fall of Einstein's Hole Argument", manuscript, pp. 1-15.

Stalzer, Theodore, *Beyond Einstein: a re-interpretation of Newtonian dynamics*, 1936, Philadelphia: Dorrance and company. (Einstein's personal library, the Einstein Archives)

Stern, Fritz, *Einstein's German World*, 1999, Princeton University Press, Princeton.

Torretti, Roberto, *Relativity and Geometry*, 1983/1996, Ney-York: Dover.

Varičak, Vladimir, "Zum Ehrenfestschen Paradoxon", *Physikalische Zeitschrift* **12** (1911), p. 169.

Winteler-Einstein Maja, *Albert Einstein –Beitrag für sein Lebensbild*, 1924, reprinted in abridged form in *The collected papers of Albert Einstein*, Vol 1, 1987, pp xlviii-lxvi. Einstein Archives, Jerusalem: the full biography printed in a typing machine with a few missing pages and double pages.

Winkelmann, Dr. A. unter Mitwirkung von R.Abegg et al., *Handbuch der Physik*, herausgegeben von. Band 4 Elektrizität und Magnetismus. (Einstein's personal library, the Einstein Archives)